\documentclass[fleqn,usenatbib]{mnras}

\usepackage{amsmath}
\usepackage[varg]{txfonts}

\usepackage[T1]{fontenc}

\DeclareRobustCommand{\VAN}[3]{#2}
\let\VANthebibliography\thebibliography
\def\thebibliography{\DeclareRobustCommand{\VAN}[3]{##3}\VANthebibliography}

\usepackage{graphicx}	
\usepackage{amsmath}	

\usepackage{natbib}
\usepackage{enumitem}
\usepackage[flushleft]{threeparttable}
\usepackage{colortbl}
\usepackage[normalem]{ulem}
\usepackage{color}
\usepackage{bm} 
\usepackage{gensymb}
\definecolor{darkblue}{rgb}{0.0,0.0,0.8}
\definecolor{darkred}{rgb}{0.75,0.,0.25}
\definecolor{darkorange}{rgb}{1,0.3,0.0}
\definecolor{darkgreen}{rgb}{0.0,0.6,0.0}
\definecolor{darkpurple}{rgb}{0.8,0.,0.9}
\definecolor{brown}{rgb}{0.65,.16,0.16}
\definecolor{grey}{rgb}{0.4,0.5,0.6}
\definecolor{white}{rgb}{1,1,1}
\definecolor{trolleygrey}{rgb}{0.5, 0.5, 0.5}
\definecolor{lavender}{rgb}{0.835,0.812,0.969}
\definecolor{pastelorange}{rgb}{0.99,0.92,0.82}
\definecolor{pastelblue}{rgb}{0.85,0.93,0.99}

\newcommand{\gaia}{{\sl Gaia}}
\newcommand{\hst}{{\sl HST}}
\newcommand{\jwst}{{\sl JWST}}

\newcommand{\msun}{\rm M_\odot}

\newcommand{\probP}{\text{I\kern-0.15em P}}
\newcommand{\diff}{{\rm d}}
\newcommand{\round}[1]{\ensuremath{\lfloor#1\rceil}}

\title[Signatures of dark matter subhalos]{Signatures of dark subhalos in dwarf spheroidal galaxies: \\ I. Fluctuations in surface density}

\author[Vitral et al.]{Eduardo Vitral$^{1}$\thanks{Email: eduardo.vitral@roe.ac.uk}\thanks{Royal Society Newton International Fellow},
Jorge Pe\~narrubia$^{1,2}$ and
Matthew G. Walker$^{3}$
\\
$^1$Institute for Astronomy, University of Edinburgh, Royal Observatory, Blackford Hill, Edinburgh EH9 3HJ, UK\\
$^2$Centre for Statistics, University of Edinburgh, School of Mathematics, Edinburgh EH9 3FD, UK\\
$^3$McWilliams centre for Cosmology and Astrophysics, Department of Physics, Carnegie Mellon University, Pittsburgh, PA 15213, USA\\
}

\date{Accepted XXX. Received YYY; in original form ZZZ}

\pubyear{\the\year{}}

\begin{document}
\label{firstpage}
\pagerange{\pageref{firstpage}--\pageref{lastpage}}
\maketitle

\begin{abstract}
Dark matter (DM) subhalos offer critical tests of cosmological models through their abundance and properties, yet most remain undetectable due to their lack of stars. We investigate whether their presence leaves measurable imprints on the projected stellar density fields of dwarf spheroidal galaxies (dSphs). Building on literature $N$-body experiments, we show that subhalo interactions induce subtle out-of-equilibrium fluctuations appearing as density corrugations. In a CDM framework, these fluctuations are dominated by the most massive subhalos in the host halo.
We develop a Fourier-based framework to quantify these features, identifying characteristic peaks in the spatial frequency spectrum that are well described by Voigt profiles. The peak parameters are sensitive to both the subhalo mass function and the number of stellar tracers. For the configurations tested, $N_{\star} \sim 10^5$ stars suffice to detect subhalo populations with $M_{\rm subhalo} \lesssim 10^6~\msun$, while larger masses produce stronger and more complex signatures.
We assess the feasibility of this technique by analyzing \gaia\ and \hst\ data: 
in this context, the Fornax dwarf shows residual low-frequency structures resembling those in our controlled subhalo experiments, making it an interesting case for follow-up.
Prospectively, wide-field surveys such as \textit{Euclid}, the \textit{Nancy Grace Roman Space Telescope}, and the \textit{Vera C. Rubin Observatory} are expected to deliver stellar samples of $N_{\star} \sim 10^5$ per dwarf, offering compelling prospects for probing subhalo imprints. Our results introduce a novel pathway to constrain the subhalo mass function in dSphs, and motivate follow-up work that incorporates alternative DM models and additional dynamical perturbations.
\end{abstract}



\section{Introduction} \label{sec: intro}

Observations across many cosmic scales suggest the presence of a dominant, non-luminous matter component, known as dark matter (DM), that governs the formation and dynamics of galaxies and clusters \citep*{Miller&Prendergast68,Hockney&Hohl69,Hohl71,Ostriker&Peebles73,Ostriker+74,Faber&Gallagher79,Frenk&White12}. Despite its success in reproducing numerous cosmological observables, the standard cosmological framework, $\Lambda$ Cold Dark Matter ($\Lambda$CDM), faces open challenges and tensions \citep{Pawlowski18,Boldrini21,Perivolaropoulos&Skara22}. These issues, together with the unknown nature of DM, have motivated both alternative dark matter scenarios \citep*{Nadler+23,Luu+25} and modifications of Newtonian dynamics at low accelerations \citep{Milgrom83,Famaey&McGaugh12,Banik&Zhao22}, each with their own successes and difficulties \citep[e.g.][]{Pointecouteau&Silk05,Tulin&Yu18}. With such caveats in mind, we focus hereafter in the standard cosmological picture, where galaxies are embedded in extended DM halos that host a hierarchy of smaller, gravitationally bound subhalos \citep{Zavala&Frenk19}.

These subhalos naturally arise from structure formation and are key to testing fundamental predictions of different DM models \citep{Bullock&Boylan-Kolchin17}.
The abundance, mass spectrum, and internal properties of subhalos provide crucial insight into the small-scale nature of DM. Cold, warm, and self-interacting DM models each predict different subhalo survival and disruption rates \citep{Chiang+25}, spatial distributions within host halos \citep*{Yang+23}, and internal structural properties \citep{Colin+02}. However, a major challenge arises from the fact that most of these subhalos are expected to be entirely dark, lacking the baryonic content necessary for direct detection via electromagnetic observations \citep{Benitez-Llambay&Frenk20}. This renders their identification difficult and represents a major obstacle in advancing our understanding of both DM and cosmology.

To overcome this, a range of indirect methods has been developed to probe the subhalo population in the Universe. These span a variety of techniques, including strong lensing \citep{Nadler+21,Ballard+24,Enzi+25,Cao+25,Tajalli+25}, flux anomalies in multiple imaged quasars \citep{Mao&Schneider98}, perturbations in stellar streams \citep{Erkal+16}, wakes in a partially ionized interstellar medium \citep{Delos25}, and high-energy emission signatures \citep{Zechlin+12}. 
While each of these techniques can in principle provide important constraints, they remain limited by factors such as sensitivity to subhalo mass, spatial coverage, and dependence on complex modeling. Although some studies suggest the potential to probe subhalos down to masses of $\sim 10^{5}\msun$ \citep[e.g.][]{Bovy+17, Drlica-Wagner+19}, claimed detections to date are generally limited to significantly higher masses, around $\sim 10^{7}\msun$ \citep[e.g.][]{Bonaca+19, Banik+21}.
This remains significantly above the cutoff masses predicted by many DM models,\footnote{While Cold Dark Matter (CDM) halos may theoretically form with masses as low as $10^{-6}~\msun$ \citep{BFPR84, Wang+20}, thus comparable to planetary scales, there is ongoing debate about the survivability of such low-mass structures in the presence of tidal disruption and other environmental effects \citep{Taylor&Babul01, Hayashi+03, Errani&Navarro21}.} leaving a key portion of the subhalo mass function currently unconstrained.

Dwarf spheroidal galaxies (dSphs) offer a promising alternative environment to investigate DM substructure. Their high DM content \citep{Pryor&Kormendy&Kormendy90}, minimal baryonic feedback \citep{Bullock&Boylan-Kolchin17}, and relatively simple stellar populations and evolutionary histories (\citealt*{Tolstoy+09}; \citealt{Savino+25}) make them ideal laboratories for isolating the effects of dark subhalos \citep[e.g.][]{Penarrubia+24}. Recently, \citeauthor{Penarrubia+25} (\citeyear{Penarrubia+25}, hereafter P25) performed a suite of $N$-body simulations to investigate the dynamical influence of dark subhalos on the stellar component of dSphs. These simulations revealed that subhalo interactions induce a gradual, self-similar expansion of the stellar distribution via stochastic energy injection. While the stellar density profile evolves self-similarly,\footnote{That is, its shape remains well described by a global model, despite undergoing expansion.} the velocity dispersion profiles experiences significant evolution, heating in the central regions and cooling in the outskirts -- which is reminiscent of an inverted gravothermal collapse.

In this work, we build upon the numerical data from P25 to investigate whether subhalo-induced dynamical heating leaves detectable imprints in the projected stellar density field of dSphs. Specifically, we ask whether local fluctuations of pixelated density maps can serve as observable tracers of dark subhalos. We also evaluate the prospects of detecting such features using current and forthcoming photometric surveys, aiming to connect theoretical predictions with observational feasibility.

The remainder of this paper is organized as follows. Section~\ref{sec: data} summarizes the key properties of the P25 simulations, outlining their relevant assumptions and caveats. Section~\ref{sec: methods} presents our methodology for extracting and quantifying subhalo-induced density fluctuations. Section~\ref{sec: results} describes the main outcomes of this analysis. In Section~\ref{sec: discussion}, we assess the implications of our findings in light of current and future survey capabilities. Finally, Section~\ref{sec: conclusion} summarizes our conclusions.

\section{Numerical models} \label{sec: data}

\begin{figure}
\centering
\includegraphics[width=\hsize]{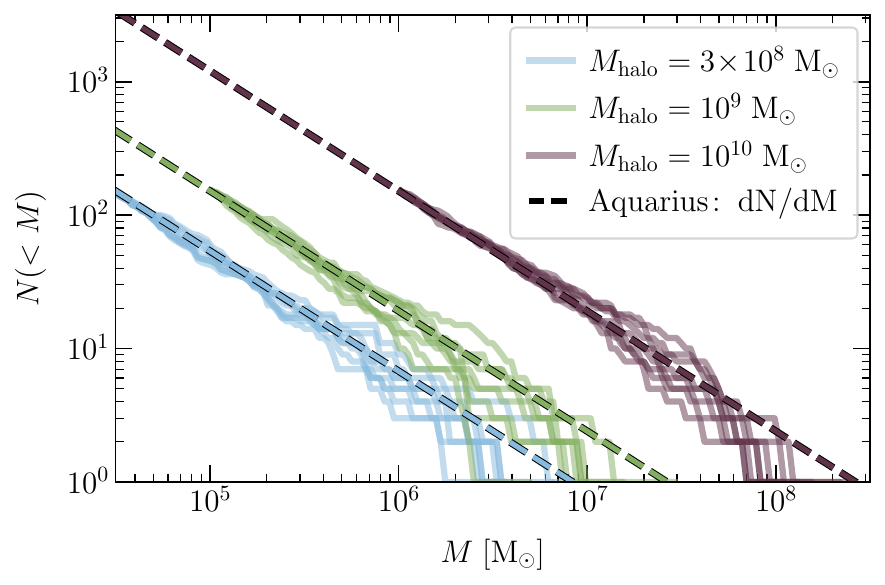}
\caption{\textit{Subhalo masses:} Cumulative number of subhalo masses for the three DM halo models considered in this study, with $M_{\rm halo}~[\msun] = \{ 3~\times~10^{8}, 10^{9}, 10^{10} \}$. 
The subhalo sampling procedure and its dependence on the host halo properties are described in section~2.3 of \protect\cite{Penarrubia+25}.}
\label{fig: subhalo-masses}
\end{figure}

\begin{figure*}
\centering
\includegraphics[width=\hsize]{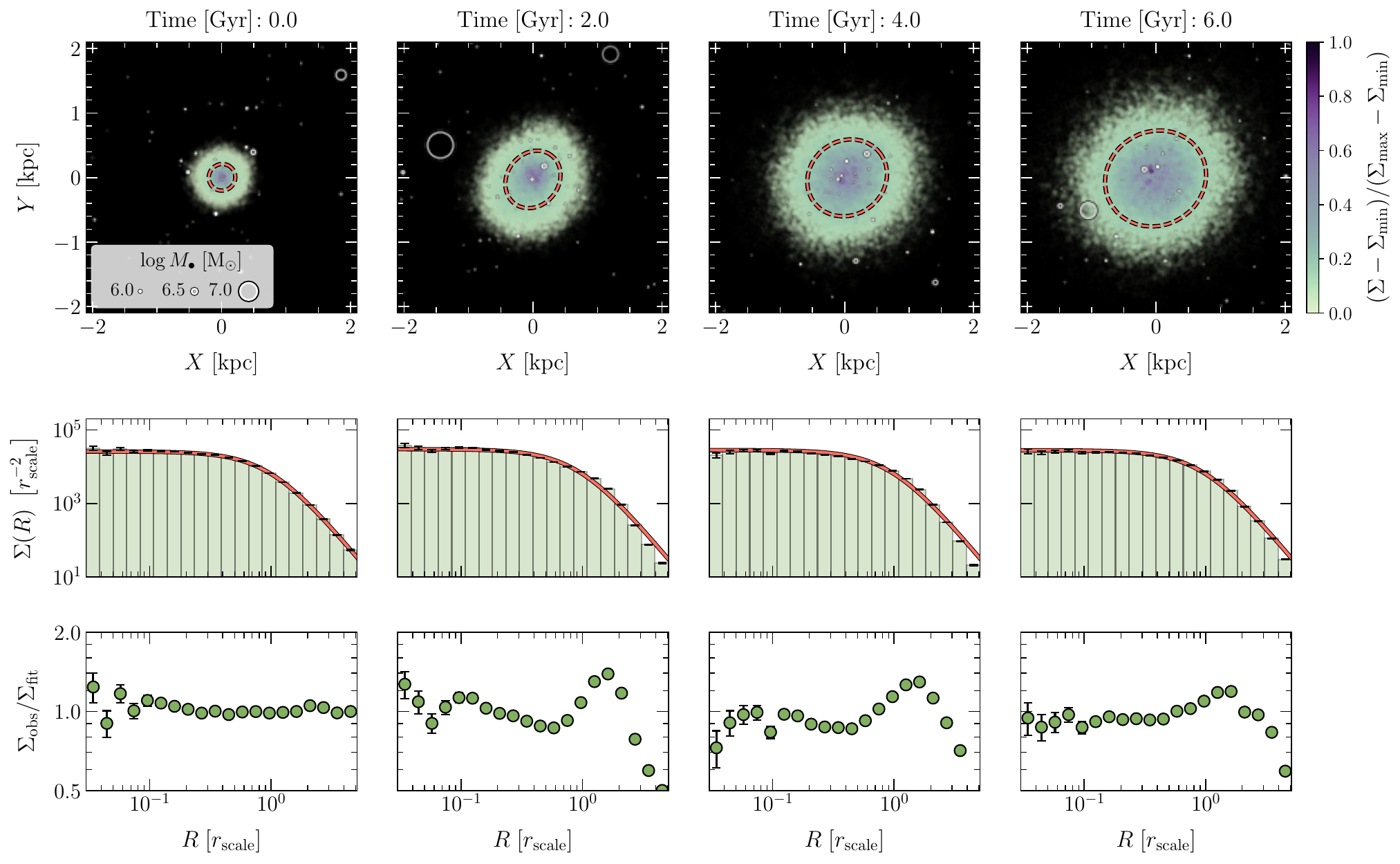}
\caption{\textit{Evolution of the Plummer distribution.} Stellar density profiles illustrating the dynamical impact of subhalos on the stellar component of the dwarf galaxy, shown at intervals of 2~Gyr per column, for a fiducial model with $M_{\rm halo} = 10^{9}~\msun$ and $N_{\star} = 10^{5}$ stellar particles. \textbf{Top row:} projected Cartesian density maps, where color and transparency scale with local relative stellar surface density; the red dashed ellipse marks the fitted projected half‐number ellipse; subhalos are plotted as white circles, their radii proportional to their log-masses and their transparency increasing with the distance from the dwarf's centre. \textbf{Middle row:} corresponding radial surface density profiles (green) versus projected distance from the galaxy centre (in unities of scale radii, defined in Eq.~\ref{eq: axi-sd-plummer}), overlaid with the best‐fitting axisymmetric Plummer model (red). \textbf{Bottom row:} radial residuals of the fit (i.e. ratio of observed and fitted $\Sigma(R)$ profiles), revealing a faint, corrugated pattern.
This figure illustrates that, to first order, an axisymmetric Plummer model provides an excellent fit to the numerical data at all stages of the dwarf's evolution.}
\label{fig: plummer-evolution}
\end{figure*}

In this work, we employ the $N$-body experiments from P25, which are designed to reproduce the effects of DM subhalos in dSphs under a set of simplifying assumptions. The construction of these models is described in detail in the original study (see their section~2), although we briefly recall a few key aspects below, for context.


The subhalo models employed in this work assume that dSphs are embedded in static DM halos that follow a \citet{Hernquist90} density profile with total masses varied as
$M_{\rm halo}~[\msun] = \{ 3 \times 10^{8}, 10^{9}, 10^{10} \}$. These masses are chosen to mimic the typical mass range expected for Local Group dwarf galaxies \citep[e.g.][]{Strigari+07,Penarrubia+08,Kravtsov10,Errani+18}. For each halo mass case, the DM scale radius is calibrated to reproduce the mean density of Milky Way dSphs with measured velocity dispersions.

The stellar component is initialized with a spherical \citet{Plummer1911} profile and $N_{\star} = 10^{5}$ particles\footnote{The impact of particle number is assessed in Section~\ref{ssec: impact-stellar-counts}, and its feasibility based on observational data is discussed in Section~\ref{ssec: expectations-stellar-counts}.}  treated as a massless tracer of the potential -- a reasonable approximation given the high mass-to-light ratios observed in dSph systems \citep[][and references therein]{Simon19}. The projected half-light radius of the stellar component is $R_{\rm half} = r_{\rm scale}$, where $r_{\rm scale}$ is the Plummer scale radius measured along the major axis.\footnote{This is independent of the chosen axis under spherical symmetry, but we adopt this notation for consistency with the remainder of this work.}
For simplicity, we set the initial scale radius to that of the Sculptor dwarf in all models, namely $r_{\rm scale} = 276$~pc.
This calibration is performed using the virial mass estimator proposed by \citet{Errani+18}, yielding halo scale radii of $r_{\rm halo}~[{\rm kpc}] = \{ 0.75, 2.26, 9.95 \}$ for the respective halo masses. We emphasize that the results presented below are largely independent of the choice of initial scale radius (P25).

To model subhalo populations, we scale the subhalo mass function from the Aquarius simulation \citep{Springel+08} to each host halo mass. The subhalo number density is expressed as a separable function of mass and radius, adopting a power-law mass function $\diff N/\diff M \propto M^{-\alpha}$ with $\alpha = 1.9$, with limits $\xi_1=M_{\rm min}/M_{\rm halo}=10^{-4}$ and $\xi_2=M_{\rm max}/M_{\rm halo}=0.03$, and a spatial distribution that follows the Hernquist profile of the host halo. For this mass interval, we obtain an average  number of $N \approx 150$ subhalos in the host halo.\footnote{As demonstrated in P25, the dominant contribution to the dynamical changes on the stellar component arises from the more massive subhalos with $\xi\approx \xi_2$. Consequently, decreasing $\xi_1$ arbitrarily increases the modeling complexity -- due to a dramatic rise in the number of subhalos -- without providing additional physical insight or improving the quality of the simulations.} Subhalos are assigned density profiles corresponding to exponentially truncated \citet*{Navarro+97}-like cusps (\citealt{Errani&Navarro21}, see also eq.~[10] of P25), with scale radii determined by the mean density of the host halo at each subhalo's pericentric radius \citep{Errani&Navarro21,Aguirre-Santaella+23}. This yields a well-defined mass–size relation, approximately following $r \propto M^{\eta}$ with $\eta \approx 0.38$. The resulting subhalo mass distributions for each DM halo model are shown in Figure~\ref{fig: subhalo-masses}.

To compute the orbits of subhalos, we treat them as non-interacting test particles moving in the static host halo potential. The initial velocities are drawn from an \cite{Osipkov79,Merritt85_df} distribution function chosen to match cosmological expectations for orbital anisotropy. Hence, the anisotropy radius is set to $r_{\rm a} = 0.3 \, r_{\rm halo}$ \citep{He+24}, ensuring approximately isotropic motions in the stellar-dominated inner regions and radially anisotropic orbits at larger radii \citep[e.g.][]{Orkney+23}. The adopted framework thus captures the key statistical and dynamical properties of cosmologically motivated subhalo populations in DM halos on the mass-scale of dSphs.


\begin{figure*}
\centering
\includegraphics[width=\hsize]{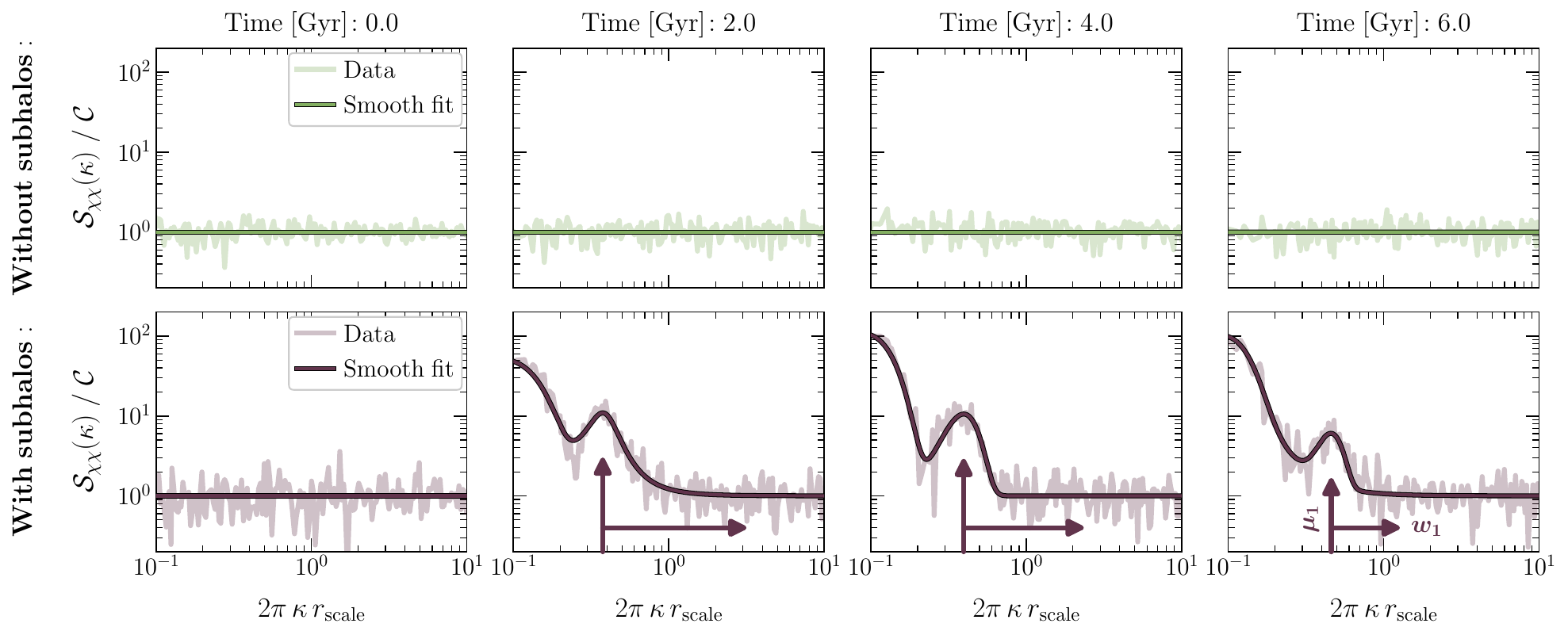}
\caption{\textit{Evolution of the power spectrum:}
Azimuthally averaged power spectrum of the $\chi$ statistic (Eq.~\ref{eq: chi-measure-poisson}) normalized by the high-frequency floor $\mathcal{C}$ (Eq.~\ref{eq: power-spectrum}), plotted against the normalized radial spatial frequency, $2 \pi \, \kappa \, r_{\rm scale}$ (see Eq.~\ref{eq: axi-sd-plummer} for the definition of $r_{\rm scale}$). Each column corresponds to a different evolutionary stage of the simulation. The top row shows equilibrium mock spectra based on the snapshot's best-fit Plummer profile, while the bottom row displays spectra from the corresponding run with subhalos. Transparent curves trace the raw signal, while solid lines show the Voigt fits using Eq.~\ref{eq: power-spectrum}. Where a significant feature is present, the primary peak defines the fitted position $\mu_{1}$ and width $w_{1}$ of the main signal, while secondary peaks at lower frequencies are attributed to smoothing artifacts (see Appendix~\ref{app: chi-reconstruct} for details and interpretation). The fiducial simulation contains $N_{\star}=10^{5}$ stellar particles in a halo of mass $M_{\rm halo}=10^{9}~\msun$. The emergence of Voigt components at $\kappa < 1$ -- absent in the control runs -- is a robust signature of subhalo–dwarf spheroidal interactions, made possible by the controlled nature of the simulations, which exclude additional perturbations such as tidal forces or major mergers.}
\label{fig: power-spectrum}
\end{figure*}

\section{Methods} \label{sec: methods}

\begin{figure*}
\centering
\includegraphics[width=\hsize]{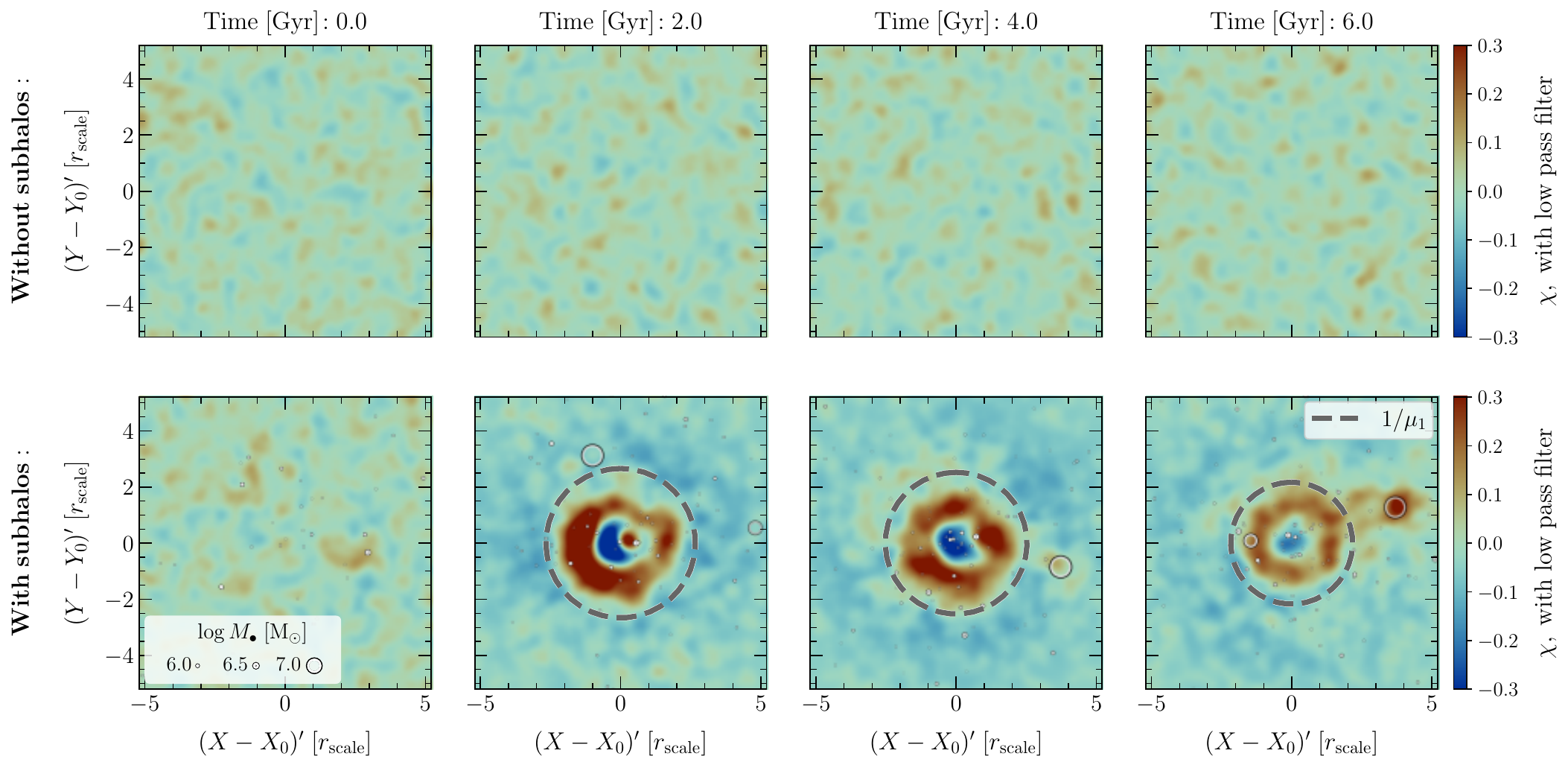}
\caption{\textit{Low-pass filter:}  Low-pass filtered $\chi$ fields for the same simulation snapshots shown in Figure~\ref{fig: power-spectrum}, constructed using Eqs.~\ref{eq: low-pass} and \ref{eq: butterworth}. The maps are projected in a rotated and centered reference frame aligned with the galaxy’s photometric axes, and spatial coordinates are normalized by the instantaneous scale radii, $r_{\rm scale}$, as defined in Eq.~\ref{eq: axi-sd-plummer}. When present, subhalos are plotted as white circles, their radii proportional to their log-masses and their transparency increasing with the distance from the dwarf's centre. All panels share a consistent color scale for visual clarity. These filtered maps highlight coherent structures resulting from subhalo interactions, including central stellar depletion and ring-like accumulations at larger radii. We display a dashed gray circle to mark the characteristic scale of such ring-like oscillations, located at $1/\mu_{1} \approx 2.5 \, r_{\rm scale}$. In some cases, localized overdensities also emerge, tracing stars temporarily captured by individual subhalos -- such as the feature near $((X - X_0)', (Y - Y_0)') \approx (4, 2) \, r_{\rm scale} \times r_{\rm scale}$, in the lower rightmost panel.}
\label{fig: low-pass}
\end{figure*}

The dominant dynamical effect identified in P25 is the gravothermal expansion of the stellar distribution, driven by the fluctuating gravitational potential sourced by subhalos. This behavior is illustrated in the top panels of Figure~\ref{fig: plummer-evolution}. 
P25 showed that the expanding stellar component remains well described by a spherical Plummer profile throughout the system's evolution. Here, we show that this result also holds for an axisymmetric Plummer distribution. To that end, we recover the best-fitting axisymmetric Plummer model at each simulation snapshot using the formalism outlined in Appendix~\ref{app: axi-plummer-fit}. The fit quality is shown in the middle panels of Figure~\ref{fig: plummer-evolution}, along with the corresponding half-number ellipse and fit residuals, respectively, in the upper and bottom panels. 
The persistent agreement between the numerical data and a Plummer profile, regardless of the dwarf’s evolutionary stage, is a central result from P25, which we adopt throughout this work. This does not imply, however, that our conclusions apply only within this specific stellar spatial distribution regime: the impact of alternative baryonic density profiles is discussed in detail later, in Section~\ref{sssec: baseline}.

While an axisymmetric Plummer profile provides a good description of the spatial distribution of stars at leading order regardless of the dwarf’s evolutionary stage, we identify second-order signatures of dSph–subhalo interactions within a galaxy’s stellar density profile. In particular, Figure~\ref{fig: plummer-evolution} suggests that one must look beyond changes in the global surface density parametrization. Therefore, we expect long-term effects such as the corrugations seen in the Figure's bottom panels to manifest as subtle, higher-order deviations from the smooth, baseline Plummer model.

To capture these deviations, we develop a statistical framework designed to quantify local fluctuations in stellar number counts at a given snapshot. We begin by overlaying a grid on the projected stellar distribution\footnote{See Appendix~\ref{app: count-cell} for technical details regarding the grid construction.} and measuring the number of stars within each cell. These counts are then compared to expectations derived from an equilibrium configuration, using the following $\chi$ statistic:
\begin{equation} \label{eq: chi-measure}
\chi \equiv \frac{N_{\rm obs} - \left<N\right>}{\sqrt{\left<N^{2}\right> - \left<N\right>^{2}}} \ ,
\end{equation}
where $N_{\rm obs}$ is the observed stellar count in a given cell, and $N$ is a random variable representing the expected distribution of counts under equilibrium conditions.

Two key assumptions underpin this analysis. First, we assume that the stellar counts in each cell follow a Poisson distribution in equilibrium -- an approximation commonly used in astrophysical contexts \citep[e.g.][]{Bahcall&Soneira80}. This implies that we neglect potential numerical effects such as softening artifacts or particle number truncation, which could induce deviations from a purely Poissonian behavior.\footnote{In real observational data, additional deviations can arise from contaminants such as foreground stars and background galaxies, which act as interlopers in the object counts. These can be addressed either by incorporating them into the expected stellar density model or by attempting to remove them during data preprocessing.} Second, following the results of P25 and the trends shown in Figure~\ref{fig: plummer-evolution}, we assume that the stellar density field is, to first order, well described by an axisymmetric Plummer model at all times, although this method can be easily applied to other stellar profiles.

Using Poisson statistics, the probability of observing $n$ stars in a cell is
\begin{equation}
\probP(N = n) = \frac{N_{\rm P}^{n} \, e^{-N_{\rm P}}}{n!} \ ,
\end{equation}
where $N_{\rm P}$ is the expected stellar count in each cell, based on a Plummer model fitted independently at each simulation snapshot (see Appendix~\ref{app: count-cell} for the analytical formalism used to compute this quantity). In these fits, the scale radius, flattening, position angle, and centroid of the stellar distribution are all treated as free parameters, ensuring that $N_{\rm P}$ consistently reflects the best-fit model at each time step. Next, the Poissonian framework allows us to express the $\chi$ statistic in a simplified form:
\begin{equation} \label{eq: chi-measure-poisson}
\chi = \frac{N_{\rm obs} - N_{\rm P}}{\sqrt{N_{\rm P}}} \ .
\end{equation}

To better understand and quantify this $\chi$ measure, we analyze its distribution in Fourier space, given the multiple advantages this domain offers. First, it enables the identification of coherent, large-scale structures that are not easily distinguishable in real space due to noise or overlapping small-scale features (see, e.g., Figure~\ref{fig: filtered-chi} in Appendix~\ref{app: chi-reconstruct} for the original $\chi$ field). Second, the Fourier basis allows for natural and controllable noise filtering, enhancing signal clarity without arbitrary smoothing kernels. Unlike model-dependent techniques, such as fitting a mixture of density components, this approach offers a model-agnostic reconstruction of the full fluctuation map, capturing the spatial scales most relevant to subhalo-induced perturbations.
This strategy is also conceptually aligned with cosmological analyses, where similar tools are used to extract physical information from temperature fluctuations in the cosmic microwave background and constrain characteristic spatial scales imprinted by early-universe processes \citep[e.g.][]{Bond&Efstathiou84,Plank20-powerspectra}.

To implement this, we construct the discrete Fourier transform of $\chi$, for each grid cell, using the \texttt{fft.fft2} module from \textsc{NumPy} \citep{vanderWalt11}, generating a map in frequency space. The spatial frequencies $k_x$ and $k_y$ are sampled with the \texttt{fft.fftfreq} module, based on the input grid window length and sample spacing. To maintain consistency across different dSph models, we normalize the sample spacing by the galaxy’s scale radius, such that the resulting frequencies $k_{i}$ are expressed in units of cycles per $r_{\rm scale}$, with $r_{\rm scale}$ updated at each snapshot. This means that the exponent in the Fourier integral takes the form $i 2\pi \omega t$, rather than the angular frequency convention, $i \omega t$, sometimes adopted in the literature.\footnote{This non-angular convention follows the original formulation by \cite{Cooley&Tukey65}, which is adopted as the default in the \texttt{fft} module \\(\url{https://numpy.org/doc/stable/reference/routines.fft.html}), while $\omega$ and $t$ denote frequency and time in a general Fourier context.}

With the Fourier transform $\widehat{\chi}\equiv {\rm FT}[\chi]$ in hand, we analyze its dependence on the radial spatial frequency, defined as $\kappa \equiv \sqrt{k_{x}^{2} + k_{y}^{2}}$, where $\kappa$ is not to be confused with the two-dimensional frequency vector $\textbf{k} = (k_{x}, k_{y})$. 
To this purpose, we compute the azimuthally averaged power spectrum, $\int \left[ \widehat{\chi} \ \widehat{\chi}^{*} \right](\kappa, \phi) \, \diff \phi$. While full reconstruction of the $\chi$ field requires phase information, this one-dimensional handling offers a clearer view of the dominant spatial modes at the scales considered, though some localized features may be missed.

As a benchmark for comparison, we generate a mock stellar distribution drawn from an equilibrium axisymmetric Plummer profile using the best-fitting parameters from each snapshot. This mock system contains the same number of particles as the simulation, providing a consistent baseline for comparison. Applying the same Fourier analysis to this control dataset allows us to isolate which features in the fluctuation spectrum are specifically attributable to the dynamical influence of subhalos.

\section{Results} \label{sec: results}

Using the formalism developed in Section~\ref{sec: methods}, we now apply our analysis to the simulation data described in Section~\ref{sec: data}. For clarity and consistency, we adopt the simulation with $M_{\rm halo} = 10^{9}~\msun$ and $N_{\star} = 10^{5}$ as our fiducial model when presenting results in Sections~\ref{ssec: power-spectrum} and \ref{ssec: low-pass-filter}. The broader impact of varying these assumptions is then explored in Sections~\ref{ssec: impact-halo-mass} and \ref{ssec: impact-stellar-counts}.

\subsection{Power spectrum} \label{ssec: power-spectrum}

\begin{figure*}
\centering
\includegraphics[width=\hsize]{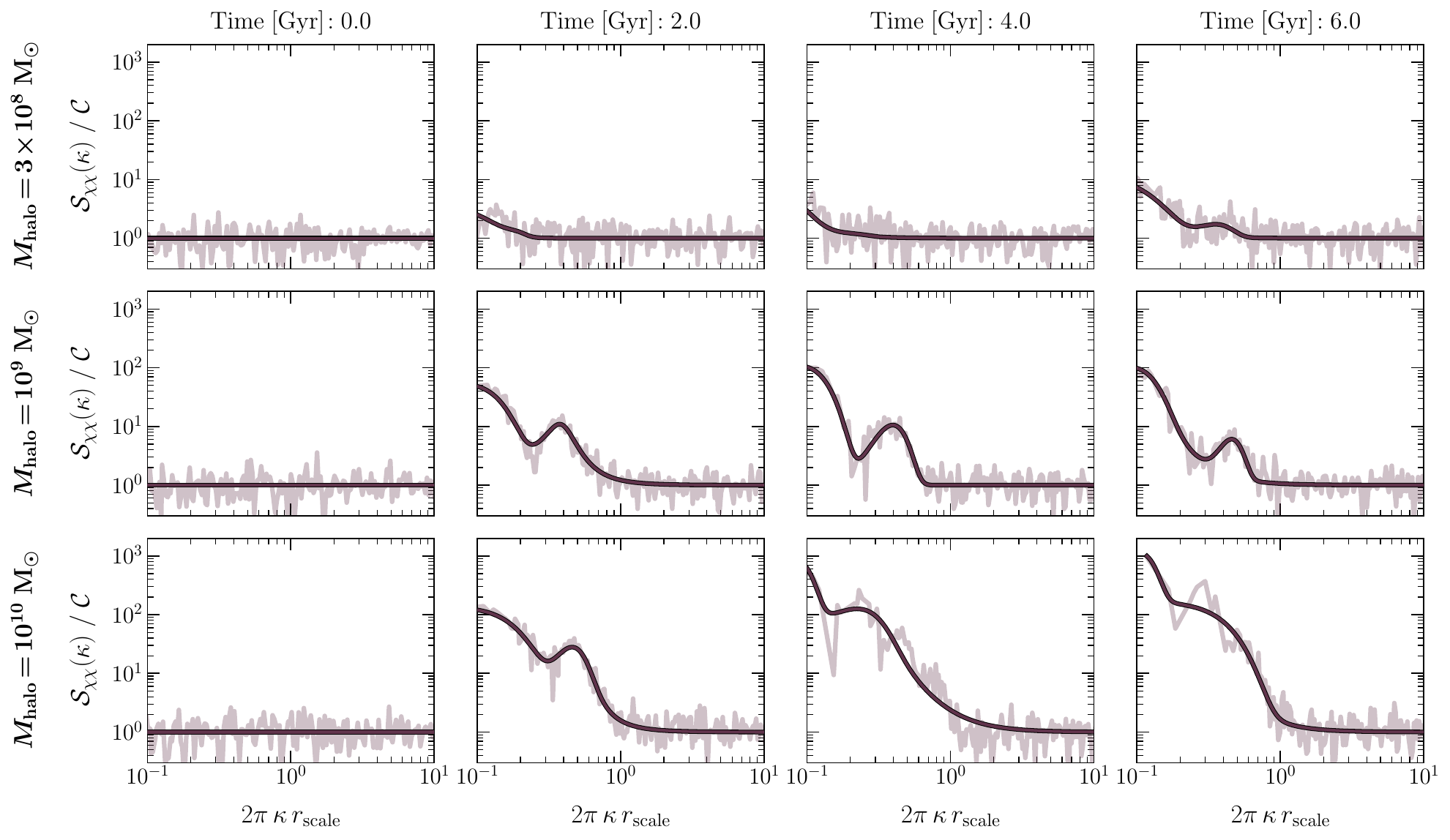}
\caption{\textit{Impact of halo mass on power spectrum:} 
Similar to Figure~\ref{fig: power-spectrum}, but here, the rows correspond to three halo–subhalo models, namely with $M_{\rm halo}~[\msun] = \{ 3 \times 10^{8}, 10^{9}, 10^{10} \}$ (top to bottom). All runs contain $N_{\star}=10^{5}$ tracer stars. The panels trace the temporal evolution of the normalized power spectrum and demonstrate its systematic dependence on the underlying mass model.}
\label{fig: mass-spectrum}
\end{figure*}

\begin{figure*}
\centering
\includegraphics[width=\hsize]{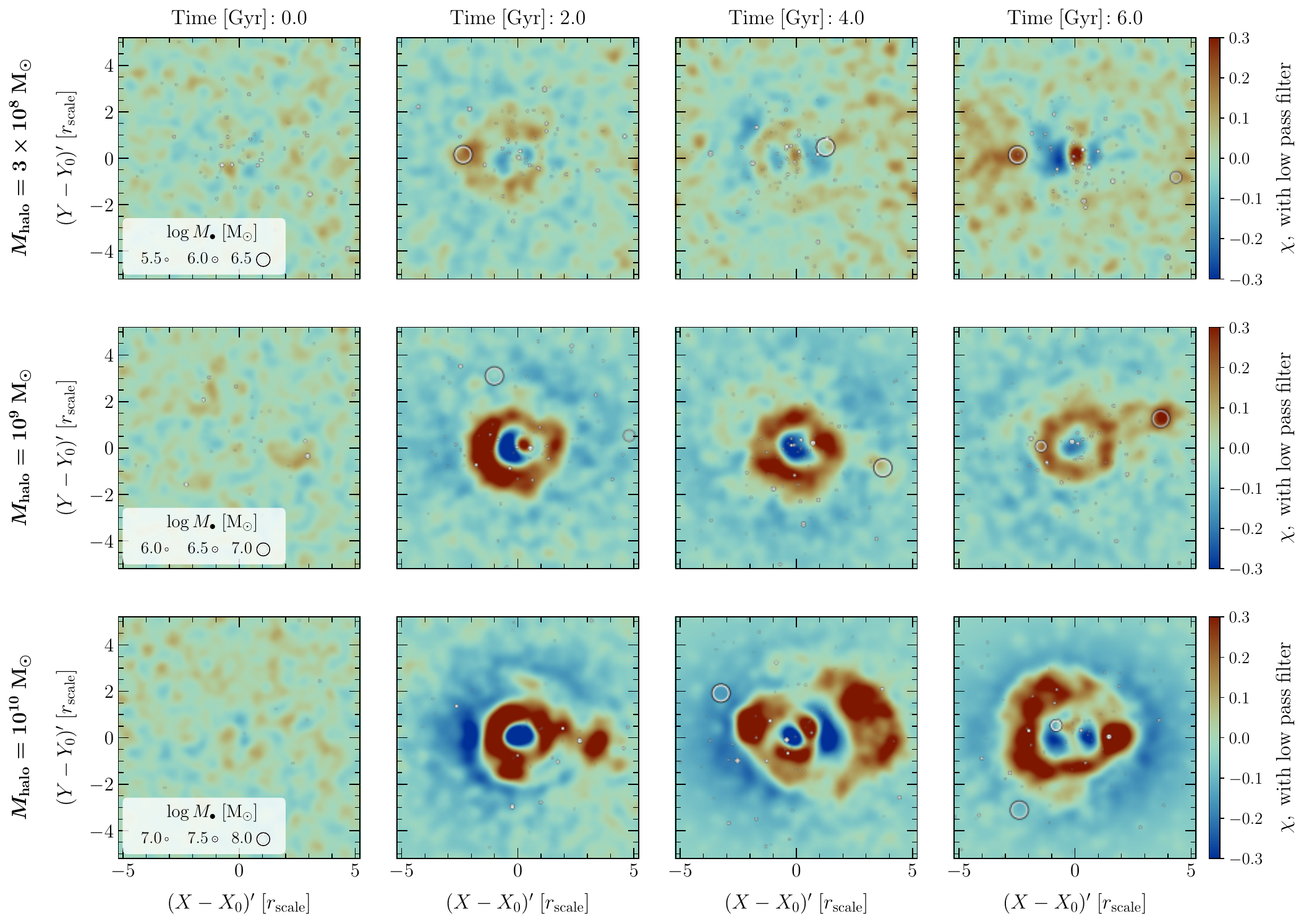}
\caption{\textit{Impact of halo mass on $\chi$ map:} 
Similar to Figure~\ref{fig: low-pass}, but here, the rows correspond to three halo-subhalo models, namely with $M_{\rm halo}~[\msun] = \{ 3 \times 10^{8}, 10^{9}, 10^{10} \}$ (top to bottom), with each row having its respective scaling of subhalo mass, for better visualization. All runs contain $N_{\star}=10^{5}$ tracer stars. The panels trace the temporal evolution of the low-pass $\chi$ map and demonstrate its systematic dependence on the underlying mass model.}
\label{fig: mass-low-pass}
\end{figure*}

Using the azimuthally averaged power spectrum, $\mathcal{S}_{\chi \chi}(\kappa) \equiv \int \left[ \widehat{\chi} \ \widehat{\chi}^{*} \right](\kappa, \phi) \, \diff \phi$,
we perform a direct comparison between a fiducial simulation that includes subhalos and a control case in equilibrium without perturbations. This comparison is shown in Figure~\ref{fig: power-spectrum}, where the top row displays the power spectrum for the equilibrium model, and the bottom row shows the corresponding deviations induced by subhalo interactions. Transparent lines represent the raw data, while the solid curves indicate smoothed fits for visual clarity. Each column corresponds to a different evolutionary stage of the simulation. For ease of comparison, the power spectra are normalized by the high-frequency noise floor $\mathcal{C}$ (see Eq.~\ref{eq: power-spectrum} below), such that the spectrum asymptotically approaches unity as $\kappa \rightarrow \infty$.

As the simulation progresses, we observe clear departures from equilibrium as subhalos begin dynamically interacting with the stellar component. These deviations become increasingly well pronounced over time, driven by the cumulative effect of repeated subhalo passages. The resulting power spectrum exhibits distinct peaks, with a first one typically located at $2 \pi \, \kappa \, r_{\rm scale} \approx 0.4$. Additional peaks at lower frequencies are also visible, which we interpret as smoothing artifacts arising from the discrete spatial sampling.\footnote{A detailed interpretation of these peaks is provided in Appendix~\ref{app: chi-reconstruct}.}

To facilitate the physical interpretation of the discrete Fourier transform, we fit a smooth functional form to the numerical spectra. We found that subhalo-induced structures in Fourier space appear atop an approximately constant high-frequency noise floor, $\mathcal{C}$, and are well described by a mixture of Voigt profiles, $V(\kappa, \mu, \sigma, \gamma)$,\footnote{Computed using the \texttt{special.voigt\_profile} module from \textsc{SciPy}.} defined as the convolution of a Gaussian (with standard deviation $\sigma$ centered at $\mu$) and a Lorentzian (with width $\gamma$). Voigt profiles are widely used when spectral features reflect multiple broadening mechanisms with distinct physical origins -- such as Doppler and pressure broadening in planetary atmospheres \citep[e.g.][]{Kuhn&London69}. In our case, this functional form offers the flexibility to accommodate both broad and narrow tails of our observed spectral peaks, which evolve across simulation snapshots (see Appendix~\ref{app: voigt-mixture} for details).

The inverse Fourier transform of a single Voigt component in 1D can be written analytically as a function of $R = \sqrt{X^{2} + Y^{2}}$ according to
\begin{equation} \label{eq: voigt_real}
\displaystyle{{\rm FT}^{-1}[ V(\kappa)]= \exp{\left(- \frac{R^2 (2\pi\sigma)^2}{2} -(2\pi\gamma) \, |R| + i \, (2\pi \mu) \, R\right)}} \ ,
\end{equation}
where the inverse Fourier transform is normalized to unity for simplicity, and the factor of $2\pi$ accounts for our use of standard (non-angular) frequencies in cycles per $r_{\rm scale}$, following the convention presented in Section~\ref{sec: methods}.
Notice that $\sigma$ and $\gamma$ control the shape of the $\chi$ distribution, which becomes a Gaussian in the limit $\gamma\to 0$, and an exponential in the limit $\sigma\to 0$. Meanwhile, the Gaussian peak in frequency space ($\mu$) sets the oscillation frequency in real space, with zeros of the real part located at $\mu R = (2n+1)/4$, where $n$ is an integer. The term $\exp(i\,2\pi \mu R)$ is a sinusoidal in real space, and therefore encodes the oscillatory pattern of alternating under- and overdensities seen in the $\chi$ maps.\footnote{The fitted frequency $\mu$ fixes the characteristic spacing of these oscillations, producing successive regions of depletion and enhancement at scales of $\approx 1/\mu$.} Larger $\mu$ values thus correspond to faster oscillations of $\chi(R)$ with radius.\footnote{A visual illustration of this functional form is provided in Figure~\ref{fig: q-vs-t}.}

We then choose to model the azimuthally averaged power spectrum as a mixture of Voigt components
\begin{equation} \label{eq: power-spectrum}
\mathcal{S}_{\chi \chi, \, {\rm fit}}(\kappa, \{\theta\}) = \mathcal{C} + \sum_{i=1}^{n} f_i \, V(\kappa, \, \mu_i, \ q_i \, \delta_i, \ (1 - q_i) \, \delta_i) \ ,
\end{equation}
where the parameter set $\{\theta\}$ includes the constant noise floor $\mathcal{C}$ and, for each peak, its flux $f_i$, central frequency $\mu_i$, characteristic scale $\delta_i$, and mixing parameter $q_i$, which regulates the relative contribution of the Gaussian and Lorentzian widths. With this notation, the Gaussian component has a standard deviation of $\sigma=q \, \delta$, while the Lorentzian component has a scale length of $\gamma=(1 - q) \, \delta$.

Because a Voigt profile blends Gaussian and Lorentzian components, its width does not follow a simple relation to either scale length individually. To provide an intuitive measure, we then adopt the approximation of \citet{Whiting68}, accurate to better than $\lesssim 1\%$ \citep{Olivero&Longbothum77}. Written in terms of the fitted parameters $\delta_{i}$ and $q_{i}$, this effective width is defined as
\begin{equation}    \label{eq: voigt-width}
    w_{i} \equiv \frac{\delta_{i}}{2} \, \left[ 1 - q_{i} + \sqrt{1 + q_{i} \, (5 \, q_{i} - 2)} \right]  \  .
\end{equation}

To avoid contamination from poorly constrained low-frequency components, we restrict our fits to $2 \pi \, \kappa \, r_{\rm scale} > 0.1$. This choice is motivated by two considerations: (i) in real data, larger spatial modes are typically inaccessible due to telescope field-of-view limitations and/or tidal interferences; and (ii) peaks at lower frequencies in our simulation are dominated by smoothing features and do not correspond to physically meaningful structures (cf. Appendix~\ref{app: chi-reconstruct}). We thus limit the number of Voigt components in Eq.~\ref{eq: power-spectrum} to $n = 2$, capturing the primary peak near $ 2 \pi \, \kappa \, r_{\rm scale} \approx 0.4$ ($i = 1$) and, when present, a blended lower frequency component associated with the first smoothing feature ($i = 2)$.
The fit quality is illustrated in Figure \ref{fig: power-spectrum}, where the smooth curves trace the Voigt profiles introduced above. For each detected signal we mark its central frequency, $\mu_{1}$, and an effective width, $w_{1}$.

Notably, for the initial snapshot of the simulation -- as well as for all equilibrium control realizations -- the recovered flux of the Voigt component is consistent with zero. This confirms that the presence of a significant Voigt signal is an direct indicator of the subhalo–dSph interactions. This conclusion is, for now, restricted to the controlled nature of our simulations, which exclude other sources of perturbation such as tides, major mergers, as well as contamination from Milky Way stars or background galaxies.

Finally, a key insight from the power spectrum analysis is that all primary peaks lie below $2 \pi \, \kappa \, r_{\rm scale} = 1$, indicating that subhalo perturbations produce density oscillations extending beyond the galaxy’s scale radius. However, we cannot rule out that experiments with a larger particle number may reveal features in the power spectrum at higher wave numbers, $2 \pi \, \kappa \, r_{\rm scale} \gtrsim 1$. 
The resolution of our models motivates a complementary treatment of the signal, introduced in the next section.

\subsection{Low-pass filter} \label{ssec: low-pass-filter}

\begin{figure*}
\centering
\includegraphics[width=\hsize]{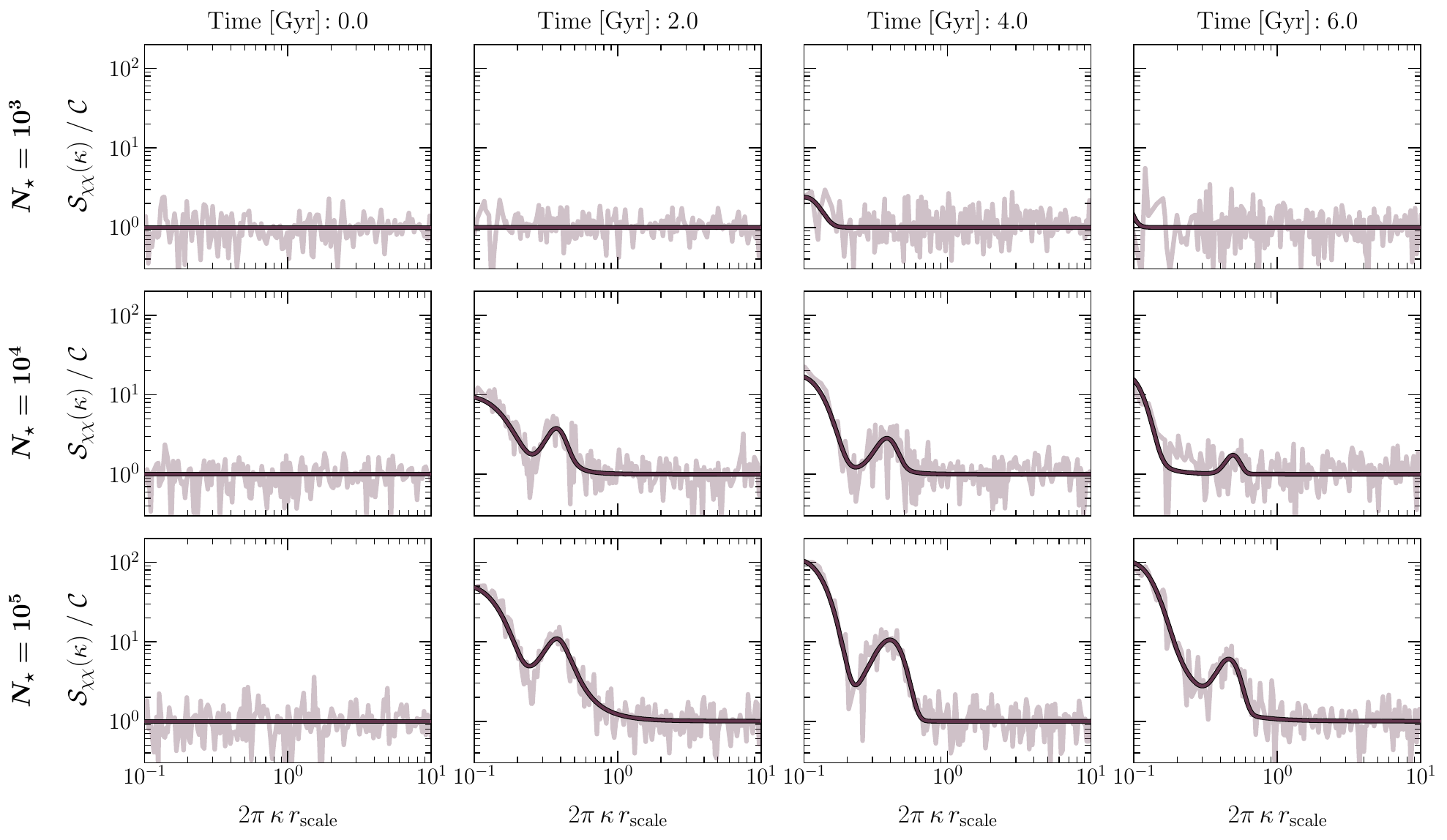}
\caption{\textit{Impact of stellar counts on power spectrum:} 
Similar to Figure~\ref{fig: power-spectrum}, but here, the rows correspond to two tracer samplings, namely with $N_{\star} = \{ 10^{3}, 10^{4}, 10^{5} \}$ particles (top to bottom). All runs pertain to the model with $M_{\rm halo}=10^{9}~\msun$. The panels trace the temporal evolution of the normalized power spectrum and demonstrate its systematic dependence on the underlying number of stellar tracers.}
\label{fig: stellar-counts-spectrum}
\end{figure*}

Given that subhalo-induced features appear predominantly in the power spectrum at frequencies $2 \pi \, \kappa \, r_{\rm scale} < 1$, a natural next step is to apply a low-pass filter to suppress higher-frequency components that may obscure the signal of interest. This allows us to isolate large-scale fluctuations driven by subhalo interactions, recovering a smoothed version of the original $\chi$ field,
\begin{equation} \label{eq: low-pass}
\chi_{\rm low~pass}(X,Y) \equiv {\rm FT}^{-1}[\widehat{\chi}(\kappa, \phi) \, B(\kappa)] \ ,
\end{equation}
with the inverse Fourier transform implemented using \textsc{NumPy}'s \texttt{fft.ifft2} module, and $B(\kappa)$ is the filtering function. To ensure a smooth transition around the cutoff frequency $\kappa_{\rm Cutoff}$, we adopt a \cite{Butterworth30} filter of the form
\begin{equation} \label{eq: butterworth}
B(\kappa) = \frac{1}{\sqrt{1 + (\kappa / \kappa_{\rm Cutoff})^{2n}}} \ ,
\end{equation}
with filter order $n = 4$ and $2 \pi \, \kappa_{\rm Cutoff} \, r_{\rm scale} = 1$, consistent with the findings from Section~\ref{ssec: power-spectrum}.

The evolution of the filtered $\chi_{\rm low~pass}$ field for our fiducial subhalo simulation is presented in Figure~\ref{fig: low-pass}, alongside the corresponding equilibrium control dataset without subhalos. As anticipated from the power spectrum, the simulation at $t = 0$ shows no significant structure beyond statistical noise, consistent with the control case. However, as the system evolves, clear spatial patterns begin to emerge. Notably, ring-like oscillations recur at characteristic intervals of $\approx 2.5~r_{\rm scale}$, consistent with the primary Voigt peaks located at $\approx 0.4$ cycles per $r_{\rm scale}$.

These features can be understood in the context of the subhalo-dSph interactions anticipated in P25. In this scenario, massive subhalos with low pericentric passages inject energy into the stellar population, pushing stars onto more extended orbits. This energy transfer drives the expansion of the stellar distribution until it asymptotically approaches the scale of the host DM halo (see P25, section~4.1). The low-pass filtered maps appear to capture corrugations that govern this expansion, revealing a spatially coherent redistribution: central regions are gradually depleted (visible as blue cores in Figure~\ref{fig: low-pass}), while stars accumulate in outer ring-like contours (highlighted in red). This form of dynamical response reflects a \textit{heating} process, where subhalo perturbations stir the stellar component across the galaxy.

Importantly, not all deviations conform to this idealized radial structure. For instance, in the rightmost panel of the lower row in Figure~\ref{fig: low-pass}, a localized overdensity appears near $((X - X_{0})^{'}, (Y - Y_{0})^{'}) \approx (4, 2)~r_{\rm scale} \times r_{\rm scale}$, tracking the path of an individual subhalo. 
Unlike the global, heating-driven corrugations, this feature is more compact and localized,\footnote{As a result, such density fluctuations are typically suppressed in the azimuthally averaged power spectrum, though they may still be visible in the full two-dimensional $(k_x, k_y)$ spectrum. In practice, their localized, phase-dependent nature makes them difficult to discern amid the noise floor in the 1D power spectrum, such that simply adding extra Voigt components to the fit would not meaningfully recover these features.} suggesting a different dynamical mechanism: stars in this region may have been temporarily displaced or even trapped in a subhalo's potential well \citep[e.g.][]{Penarrubia+24}. These localized overdensities are indicative of stellar \textit{capture} -- a complementary mode of subhalo interaction that leaves imprints distinct from the broader patterns driven by heating.

\begin{figure*}
\centering
\includegraphics[width=\hsize]{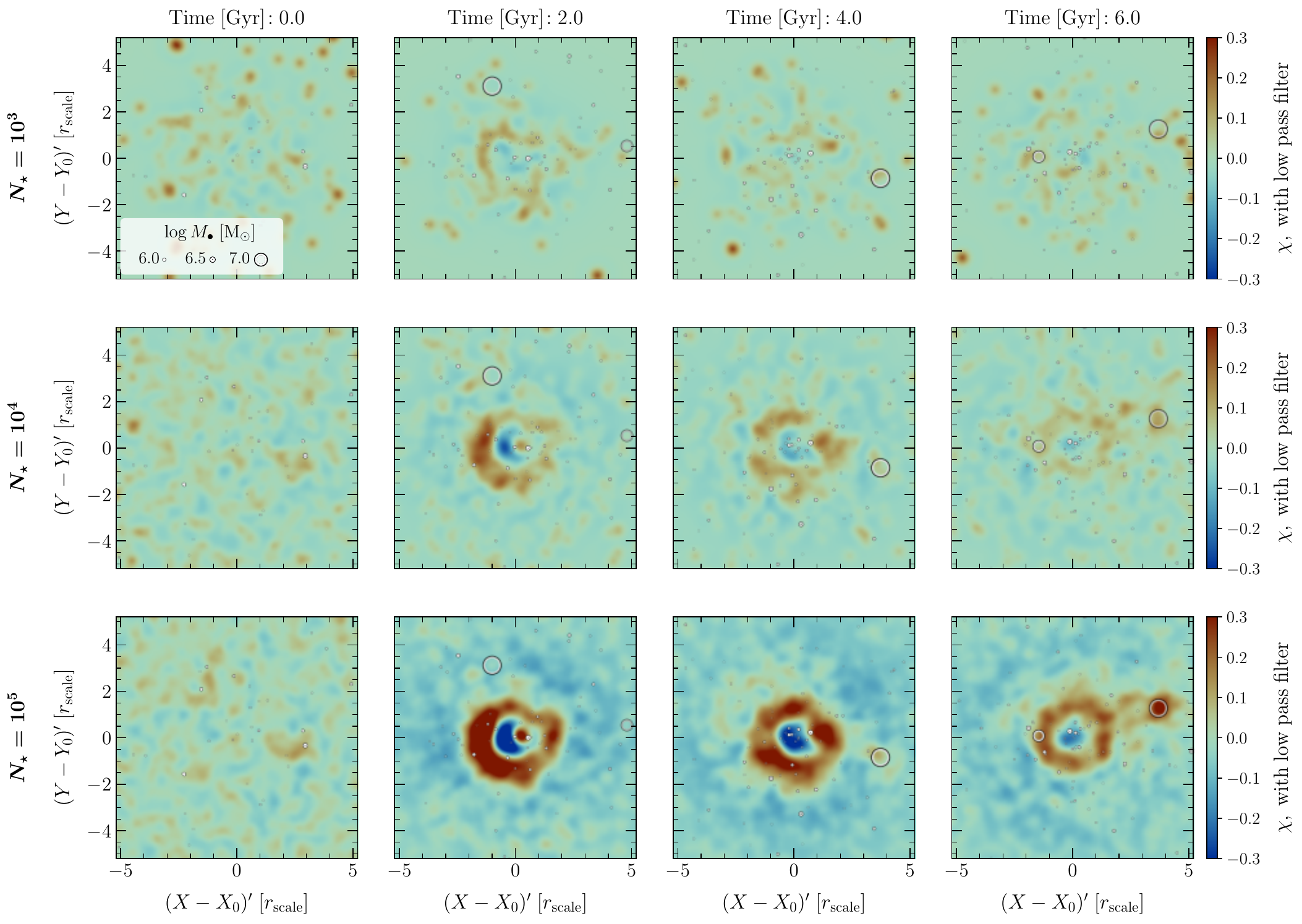}
\caption{\textit{Impact of stellar counts on $\chi$ map:} 
Similar to Figure~\ref{fig: low-pass}, but here, the rows correspond to two tracer samplings, namely with $N_{\star} = \{ 10^{3}, 10^{4}, 10^{5} \}$ particles (top to bottom). All runs pertain to the model with $M_{\rm halo}=10^{9}~\msun$. The panels trace the temporal evolution of the low-pass $\chi$ map and demonstrate its systematic dependence on the underlying number of stellar tracers.}
\label{fig: stellar-counts-low-pass}
\end{figure*}

In the remainder of the paper, we focus on the global, heating-driven signatures of subhalo interactions -- namely, the large-scale corrugations in the surface density profile revealed through the azimuthally averaged power spectrum and the filtered $\chi$ maps. This emphasis reflects both our interest in statistically characterizing the population-level impact of subhalos and the interpretability benefits of angular averaging, which simplifies the detection of dominant spatial modes while suppressing localized, phase-dependent features. A deeper exploration of stellar capture signatures in full 2D Fourier space is a promising direction for future work that may complement the heating-based approach. For now, we examine how the strength and morphology of heating-driven features depend on the DM halo mass and the number of stellar tracers. To support reproducibility and further inspection, we also provide full animations of the 1D power spectrum and the filtered $\chi$ field evolution across all simulation snapshots as online material.

\subsection{Impact of dSph halo mass}
\label{ssec: impact-halo-mass}

We now explore how the diagnostics introduced in Sections~\ref{ssec: power-spectrum} and \ref{ssec: low-pass-filter} depend on the adopted halo-subhalo mass model. Because each halo mass corresponds to a different truncation of the subhalo mass function (Figure~\ref{fig: subhalo-masses}), both the power spectrum and the low-pass $\chi$ maps can change appreciably with this quantity. Figures~\ref{fig: mass-spectrum} and \ref{fig: mass-low-pass} therefore reprise Figures~\ref{fig: power-spectrum} and \ref{fig: low-pass} for the alternative halo masses, enabling a direct, qualitative comparison across models.

Figure~\ref{fig: mass-spectrum} reveals that the normalized flux of the primary Voigt peak\footnote{In the notation of Eq.~\ref{eq: power-spectrum}, this quantity is given by $f_{1}/\mathcal{C}$.} correlates strongly with halo-subhalo mass: lower-mass models consistently yield weaker peaks, whereas higher-mass ones produce stronger signals. The same ordering appears in the low-pass $\chi$ maps of Figure~\ref{fig: mass-low-pass}: larger halo-subhalo masses generate higher-contrast spatial fluctuations relative to the baseline Plummer profile. 

A second trend is that, for most snapshots, both the primary peak position ($\mu_{1}$) and width ($w_{1}$) are larger in the $M_{\rm halo}=10^{9}~\msun$ run than in the $M_{\rm halo}=3\times10^{8}~\msun$ case. Meanwhile, the comparison with the $M_{\rm halo}=10^{10}~\msun$ model is less straightforward. At early times, the expected monotonic scaling with halo mass holds, but as the simulation evolves subhalo perturbations of stellar orbits build up. In Fourier space, the two dominant Voigt peaks blend together, reducing the precision with which Eqs.~\ref{eq: power-spectrum} and \ref{eq: voigt-width} recover $\mu_{1}$ and $w_{1}$. This complexity is echoed in Figure~\ref{fig: mass-low-pass}: the most massive halo develops increasingly complicated spatial structures -- e.g. a second outer half-annulus at 4~Gyr and multiple central depletion zones by 6~Gyr -- making its power spectrum harder to interpret. The formal uncertainties on $\mu_{1}$ and $w_{1}$ returned by the Voigt fits therefore play an important role, and can be used to assist with the quantitative distinction among different subhalo-mass scenarios -- a point we revisit in Section~\ref{ssec: subhalo-mass-func}.

\begin{figure*}
\centering
\includegraphics[width=\hsize]{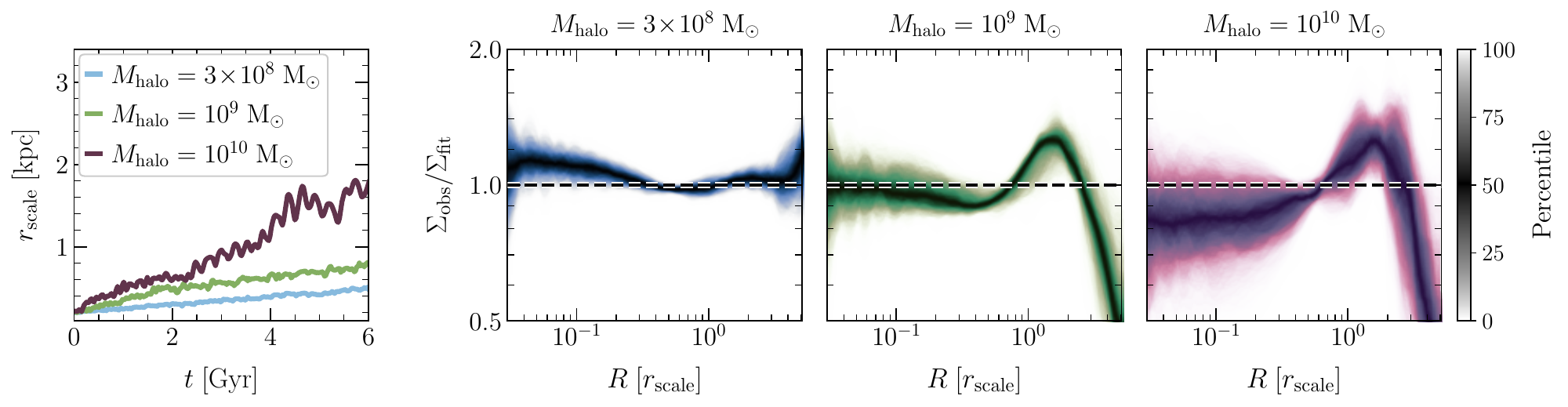}
\caption{\textit{Steady-state equilibrium:} 
Time evolution of corrugations induced by subhalo interactions in models with $N_{\star} = 10^{5}$ tracers. The left panel shows the evolution of the best-fit $r_{\rm scale}$ for three galaxy models of increasing host halo mass (blue, green, and purple curves). The remaining panels display radial surface-density ratios relative to the axisymmetric Plummer fit of each snapshot, normalized by that snapshot’s instantaneous scale radius. For each model, all snapshots are stacked and ordered by residual percentile, with shading indicating the percentile range. Subhalo interactions produce a steady-state depletion of stars inside $r_{\rm scale}$ with a corresponding accumulation beyond it. More massive subhalos amplify this pattern but also perturb it more efficiently, as seen from the broader percentile spread in the $M_{\rm halo} = 10^{10}~\msun$ panel.}
\label{fig: standing-wave}
\end{figure*}

\subsection{Impact of stellar counts}   \label{ssec: impact-stellar-counts}

To assess how our diagnostics depend on tracer sampling, we repeat the analyses of Sections~\ref{ssec: power-spectrum} and \ref{ssec: low-pass-filter} for three values of $N_{\star}$. Figures~\ref{fig: stellar-counts-spectrum} and \ref{fig: stellar-counts-low-pass} present the resulting power spectra and low-pass $\chi$ maps for a $M_{\rm halo}=10^{9}~\msun$ model. The upper row correspond to $N_{\star}=10^{3}$, middle row depict results for $N_{\star}=10^{4}$, while the lower row shows the fiducial case with $N_{\star}=10^{5}$.

The primary Voigt peak’s centre, $\mu_{1}$, and width, $w_{1}$, are somewhat consistent -- within their uncertainties -- between the two samplings. In contrast, the normalized flux $f_{1}/\mathcal{C}$ is markedly higher for $N_{\star}=10^{5}$, indicating that signal strength scales with the number of tracers. A similar trend appears in the low-pass maps: spatial trends are far more pronounced in the high-count run, though still detectable with $N_{\star}=10^{4}$, with barely no detection for $N_{\star}=10^{3}$. Adequate stellar sampling is therefore essential for an accurate characterization of subhalo-induced density fluctuations. Indeed, an additional test with $M_{\rm halo}=3\times10^{8}~\msun$ and $N_{\star}=10^{4}$ yielded no meaningful signal, such as the $N_{\star}=10^{3}$ sampling in the $M_{\rm halo}=1\times10^{9}~\msun$ model in Figure~\ref{fig: stellar-counts-low-pass}. Hence, increasing the sensitivity to low-mass subhalos requires a larger number of visible tracers. We revisit this requirement, as well as its implications for current and forthcoming surveys, in Section~\ref{ssec: expectations-stellar-counts}.

\section{Discussion} \label{sec: discussion}

\subsection{Departure of equilibrium} \label{ssec: equilibrium}

Throughout this work, we adopted an axisymmetric Plummer profile as baseline equilibrium model and sought to recover higher-order deviations induced by the dynamical influence of DM subhalos -- beyond those expected from simple Poisson fluctuations. In doing so, we identified density fluctuations within a few galactic scale radii, which we attributed to out-of-equilibrium signatures. In this section, we discuss the regime followed by such fluctuations and assess how the choice of baseline density model impacts our conclusions, using Figure~\ref{fig: standing-wave} as a visual aid.

The first, leftmost panel of this figure tracks the time evolution of the best-fit $r_{\rm scale}$ for different galaxy models (blue, green, and purple curves, from least to most massive parent halo). The remaining three panels, equivalent to the bottom row of Figure~\ref{fig: plummer-evolution}, show the radially dependent surface-density ratios,\footnote{Ratios are taken with respect to our axisymmetric Plummer profile fit.} as a function of distance to the galaxy's centre, in units of the instantaneous scale radius. Here, however, all snapshots of a given model are stacked and ordered by residual percentile to highlight how the remnant structure varies.

\subsubsection{Steady-state perturbation} \label{sssec: steady-state}

Figure~\ref{fig: standing-wave} shows that, when residuals are measured relative to a profile with a constant stellar-density core, subhalo interactions -- particularly from the more massive subhalos -- tend to produce a central depletion of stars followed by an outer accumulation\footnote{Residuals also show a steepening of the outer density profile relative to the Plummer fit, beyond a few $r_{\rm scale}$ and more pronounced in the more massive halo–subhalo models. This effect is, by construction, directly reflected in the power spectrum of the $\chi$ map. Although the outer slope is generally less constraining than the inner one (see next Section and \citealt{Ciotti&Morganti09}), the trend remains of interest, as such slopes are easier to constrain with current surveys such as \gaia.} (i.e. $\Sigma_{\rm obs} / \Sigma_{\rm fit}$ falling below and then rising above the dashed black line), in agreement with what has been presented in Section~\ref{sec: results}. The underlying mechanism starts with dark subhalos generating stochastic force fluctuations that inject energy into stellar orbits, driving a gradual gravothermal expansion of the stellar component (P25). As stars migrate outward, the inner regions ($\lesssim r_{\rm scale}$) are depleted while material accumulates around $\gtrsim r_{\rm scale}$. In the movies provided as online material, one observes that this corrugation, normalized at $r_{\rm scale}$, approaches a nearly steady-state form after a few crossing times. The consistency of the residual profiles across snapshots is supported by the narrow spread of their radial percentiles, which remain too small to alter the overall residual shape.

Still from Figure~\ref{fig: standing-wave}, we observe that more massive subhalos amplify this steady-state pattern, producing a more pronounced inner depletion and outer accumulation, but they also perturb it more efficiently. This is evident from the larger dispersion between the 0th and 100th percentiles in the $M_{\rm halo} = 10^{10}~\msun$ model (cf. rightmost panel of Figure~\ref{fig: standing-wave}). This enhanced variability underscores that, although the corrugation retains its form and resembles a steady-state wave, it is not a true dynamical equilibrium: the stars that populate it change continuously as the wave gets perturbed and its absolute scale grows through gravothermal expansion.

\subsubsection{Baseline model considerations} \label{sssec: baseline}

Our simulations provide multiple snapshots, enabling departures from equilibrium to be identified through the \textit{time variability} of the stellar density field, beyond what would be expected from Poisson fluctuations alone. For observed dSphs, by contrast, only a single snapshot is available, which makes the definition of a density baseline model more subtle. Even with extended observational baselines, the long intrinsic dynamical timescales of these galaxies preclude direct detection of significant structural changes.

One might therefore ask whether the argument for non-equilibrium becomes unnecessary if a more flexible parametrization can reproduce the observed density without leaving significant residuals. Figure~\ref{fig: standing-wave} suggests that this comes at the cost of requiring the inner density slope to be positive, i.e.
$\displaystyle \lim_{r \to 0} \left[ \diff \rho / \diff r \right] > 0$.
This implies a stellar density that decreases toward the centre, indicating an asymptotic density hole. In this light, the Plummer profile is best viewed not as a representation of an underlying equilibrium state, but as a practical tool for isolating perturbations relative to a constant-density stellar core.

Interestingly, profiles with centrally decreasing stellar density are often viewed as problematic for stability or physical plausibility (cf. \citealt{Binney&Tremaine87}, Antonov’s Fourth Law). In spherical systems, the cusp–slope anisotropy theorem of \citet{An&Evans06} shows that only certain combinations of central density slope and orbital anisotropy yield non-negative distribution functions. These conditions, originally derived for self-consistent spherical systems, have since been extended to multi-component cases, including tracers in external potentials (e.g. \citealt{Ciotti&Morganti09, Ciotti&Morganti10a, Ciotti&Morganti10b}; \citealt*{VanHese+11}). Although no formal analogue exists for an axisymmetric baryonic component embedded in a DM halo, the emergence of a centrally decreasing density remains restrictive, typically requiring weakly concentrated halos, limited or no radial anisotropy, and avoidance of the deepest regions of the potential well \citep[see][specifically]{Ciotti&Morganti09}. Curiously, figure~6 from P25 shows that our simulations maintain a radially anisotropic regime throughout the gravothermal expansion, in contrast with these requirements and thus strengthening the case for out-of-equilibrium dynamics.

\begin{figure}
\centering
\includegraphics[width=\hsize]{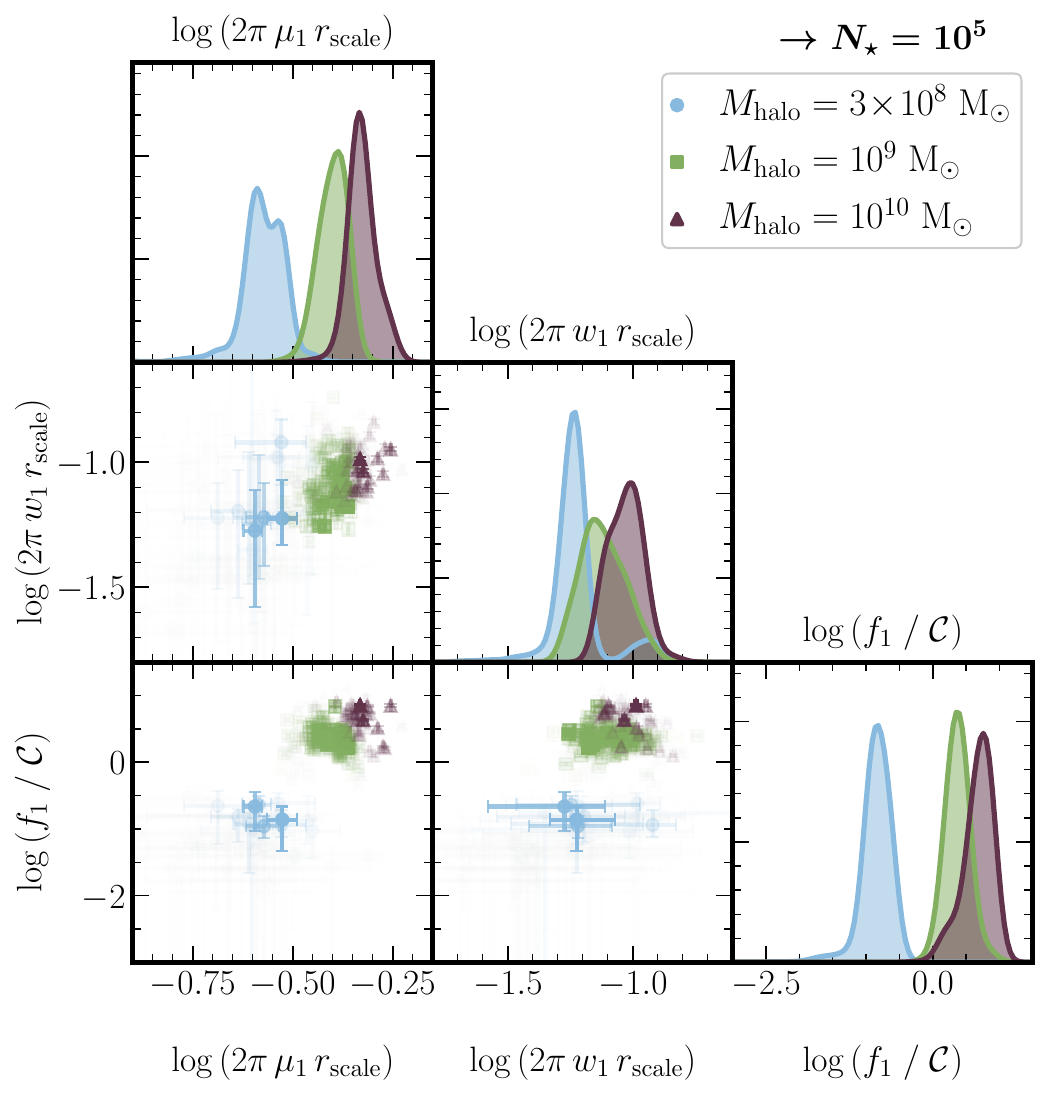}
\caption{\textit{Constraints on subhalo mass function:} 
Corner plots of the primary Voigt peak parameters from Eqs.~\ref{eq: power-spectrum} and \ref{eq: voigt-width}: central position $\mu_{1}$, width $w_{1}$, and normalized flux $f_{1} / \mathcal{C}$, all in logarithmic scale for better visualization. 
Each dataset corresponds to nearly 250 simulation snapshots with $N_{\star} = 10^{5}$ stellar particles, for halo masses of $3\times10^{8}\msun$ (blue), $10^{9}\msun$ (green), and $10^{10}\msun$ (purple). To down-weight poorly constrained fits, the one-dimensional histograms are weighted by the inverse of the square product of the parameter uncertainties $\epsilon_{\mu}$ and $\epsilon_{w}$; transparency in the two-dimensional correlation panels is scaled accordingly. The resulting distributions support the conclusion that, given sufficient stellar sampling, meaningful constraints on the underlying halo–subhalo mass model can be achievable.
}
\label{fig: corner-mass}
\end{figure}

\subsection{Prospects for subhalo mass function constraints} \label{ssec: subhalo-mass-func}

Our simulations show that subhalos imprint measurable signatures on the stellar surface density field of dSphs, manifesting both as coherent structures in the filtered $\chi$ maps (cf. Eq.~\ref{eq: chi-measure}) and as distinct peaks in the azimuthally averaged power spectrum. These differences persist across various halo–subhalo mass configurations and are well captured by Voigt mixture fits to the one-dimensional spectrum (Figure~\ref{fig: power-spectrum}). They become even clearer in the reconstructed filtered $\chi$ maps when phase information is included (Figure~\ref{fig: low-pass}).

In this section, we explore the dependence of the power spectrum on the subhalo mass function more quantitatively by analyzing the fitted Voigt parameters across all simulation snapshots. While we focus here on the azimuthally averaged power spectrum for consistency with previous sections and to facilitate interpretation of the Voigt features, the same concept can be adapted to the full two-dimensional spectrum if necessary.

Specifically, we construct corner plots for each halo mass model, using simulations with $N_{\star} = 10^{5}$ stellar particles. We compile the Voigt profile parameters derived from nearly 250 snapshot fits, focusing on those associated with the primary Voigt peak: the central position $\mu_{1}$, width $w_{1}$, and normalized flux $f_{1} / \mathcal{C}$. To reduce the influence of poorly constrained measurements, we weight the one-dimensional histograms in the corner plots by the inverse of the square product of the uncertainties $\epsilon_{\mu}$ and $\epsilon_{w}$. The transparency in the two-dimensional correlation panels is also scaled accordingly. The resulting distributions are shown in Figure~\ref{fig: corner-mass}.

These plots provide a more quantitative foundation for the conclusions drawn in Section~\ref{ssec: impact-halo-mass}, demonstrating that robust constraints on the subhalo mass function are achievable when the stellar sampling is sufficiently high. In such cases, one can statistically evaluate different halo–subhalo mass configurations by constructing a grid of models tailored to match observational data. While the configurations explored in this work represent a limited set, real observations can inform more realistic assumptions about the host halo -- e.g., through dynamical mass modeling \citep[][]{Vitral+24, Arroyo-Polonio+25}. From this starting point, one can vary the subhalo mass function parameters to match the observed power spectrum either in one or two dimensions, depending on whether phase information is retained. In this way, our framework offers a new avenue for constraining the subhalo population in the Universe through the lens Local Group dSphs.

\begin{table}
\caption{Number of stars counts vs. magnitude depth.}
\label{tab: stellar-counts}
\centering
\renewcommand{\arraystretch}{1.5}
\tabcolsep=10pt
\footnotesize
\begin{tabular}{l|rr}
\hline\hline   
\multicolumn{1}{c}{Dwarf name} &
\multicolumn{1}{c}{$G \lesssim 22$} &
\multicolumn{1}{c}{${\rm F606} \lesssim 26$} \\ 
\hline
Ursa Minor  &  $\mathcal{O}(10^{3})$  &  $\mathcal{O}(10^{5})$ \\
Draco       &  $\mathcal{O}(10^{3})$  &  $\mathcal{O}(10^{5})$ \\
Sculptor    &  $\mathcal{O}(10^{4})$  &  $\mathcal{O}(10^{5})$ \\
Fornax    &  $\mathcal{O}(10^{5})$  &  \multicolumn{1}{c}{--} \\
\hline
\end{tabular}
\parbox{\hsize}{\textit{Notes:}
Columns are: \textbf{(1)} Name of the dwarf galaxy;
\textbf{(2)} Expected number of stars within the \gaia\ magnitude range ($G \lesssim 22$);
\textbf{(3)} Expected number of stars within the \hst\ magnitude range (${\rm F606W} \lesssim 26$).
The procedure used to obtain these estimates is described in Section~\ref{ssec: expectations-stellar-counts}.
Values correspond to a field of view encompassing several half-light radii of each galaxy.
}
\end{table}

\subsection{Expectations on observable data} 

In this section, we assess the viability of our modeling approach in the context of current and upcoming astronomical surveys, and discuss key caveats to consider.

\subsubsection{Stellar counts} 
\label{ssec: expectations-stellar-counts}

As demonstrated in Section~\ref{ssec: impact-stellar-counts}, increasing the number of stellar tracers enhances the detectability of subhalo-induced signals. In this study, we adopt a fiducial stellar count of $N_{\star} = 10^{5}$, balancing the need to maximize sensitivity to dynamical perturbations while maintaining consistency with plausible observational limits.
In this section, we assess the viability of this choice in light of current and near-future observational capabilities, focusing on four classical Milky Way satellites: Ursa Minor, Draco, Sculptor, and Fornax. The first three were selected based on the availability of reduced \hst\ astrometry in our possession through the HSTPROMO collaboration,\footnote{\url{https://www.stsci.edu/~marel/hstpromo.html}} with Sculptor being the most comparable to our setup in terms of stellar density and velocity dispersion. We also include Fornax due to its excellent coverage in \gaia\ data, despite the lack of reduced \hst\ data in our sample.


For each galaxy, we explore two observational depth regimes: (i) a shallower one with an equivalent \gaia\ $G$-band magnitude limit of $\approx 22$, also representative of surveys such as SDSS or DESI; and (ii) a deeper one with an equivalent \textit{Hubble Space Telescope} (\hst) F606W magnitude limit of $\approx 26$, corresponding to the detection limits of space-based observatories such as \jwst, \textit{Euclid}, and the \textit{Nancy Grace Roman Space Telescope} (hereafter simply {\it Roman}). This deeper limit is also relevant for the \textit{Vera C. Rubin Observatory}, though observations at this depth may be adversely affected by satellite constellations \citep{Hainaut&Williams20}. 
Due to the lack of reduced \hst\ data for Fornax in our sample, we consider only the shallower \gaia-based regime for that galaxy.
In all scenarios, we assume an effectively unlimited field of view -- achievable either through instruments with large native coverage or by mosaicing smaller fields into a broader footprint. This assumption is supported by results in Section~\ref{sec: results}, which show that subhalo signatures are most prominent beyond the dwarf’s scale radius.

To estimate total stellar counts, we adopt different strategies for each magnitude regime. For case (i), we apply axisymmetric Plummer fits to \gaia-EDR3 data to determine the ratio of contaminants to member stars, following a formalism similar to \citet{Vitral21}, and extrapolate the total number of dwarf galaxy members to large radii. For case (ii), we use archival \hst\ data from \citet{2021hst..prop16737S} and \citet{Vitral+23-UMi} to count member stars within a single \hst\ field, after removing interlopers using the observed color–magnitude diagram (see Section~2.3.2 of \citealt{Vitral+24}). We then generate a mock population of one million stars from a Plummer profile fit to each galaxy and compute the fraction that falls within the \hst\ footprint. This fractional coverage is used to rescale the observed star count and recover a global estimate at the deeper magnitude limit.

The resulting order-of-magnitude estimates for the number of member stars in each scenario are summarized in Table~\ref{tab: stellar-counts}. Achieving a statistically meaningful sample of $\mathcal{O}(N) \geq 10^{5}$ generally requires the depth and field-of-view combination provided by the new generation of space-based observatories. In this regime, {\it Euclid} and {\it Roman} stand out as especially promising. Notably, Fornax constitutes an exception: due to its higher stellar density, existing \gaia\ data alone can already provide a sample of $\sim 10^5$ likely members. Thus, our adopted fiducial value of $N_{\star} = 10^{5}$ is not only motivated by the desire to maximize subhalo sensitivity, but also lies within reach of current and upcoming observatories.

\subsubsection{Application to \gaia\ data} \label{sssec: test-gaia}

Building on the arguments presented throughout this work, if Fornax hosts a DM halo with a subhalo population similar to our simulations, heating signatures from subhalo–stellar interactions should, in principle, be detectable in \gaia\ data, given its high stellar count of $N_{\star} \sim 10^{5}$. Sculptor, while less favorable with $N_{\star} \sim 10^{4}$, could still show analogous patterns at lower signal-to-noise. Naturally, such signatures are only interpretable if other dynamical processes do not produce corrugations at the same spatial frequencies. We therefore stress that the goal of this brief analysis is not to claim an indirect detection of subhalos in these systems, but rather to assess whether the specific signatures predicted by our framework could be recoverable in real data.

\begin{figure*}
\centering
\includegraphics[width=\hsize]{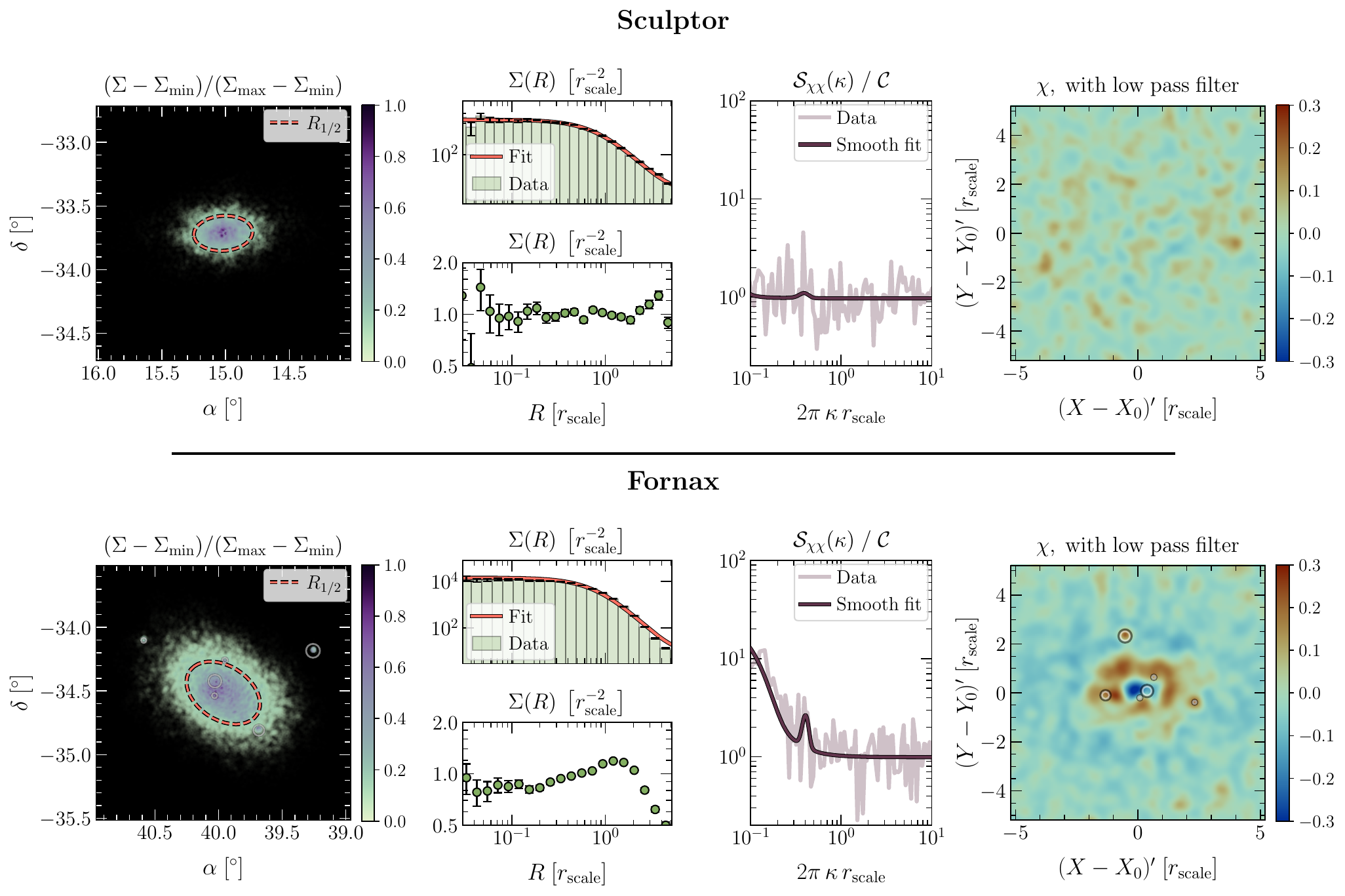}
\caption{\textit{Application to Sculptor and Fornax:}
We apply our methodology to \gaia\ DR3 data for the Sculptor (top rows) and Fornax (bottom rows) dwarf galaxies, containing an estimated $\mathcal{O}(10^{4})$ and $\mathcal{O}(10^{5})$ member stars, respectively. For each system, the \textbf{first column} shows the observed surface density map, constructed following the procedure in Figure~\ref{fig: plummer-evolution}. The \textbf{second column} replicates the two lower rows of that figure, with the addition of a constant background surface density to account for interlopers (see Appendix~\ref{app: axi-plummer-fit}). The \textbf{third and fourth} columns present the azimuthally averaged 1D power spectrum of the $\chi$ field and the corresponding low-pass filtered $\chi$ map. 
To ease comparison across panels, the radial profiles in the second column are shown over the same spatial range as the $\chi$ maps, with the background level becoming relevant beyond $\gtrsim 7\,r_{\rm scale}$ for Sculptor and $\gtrsim 5\,r_{\rm scale}$ for Fornax, respectively.
For context, the positions of Fornax’s six stellar clusters are over-plotted as white circles in the spatial maps, not to be confused with subhalos, or the stellar capture effects discussed in Section~\ref{sec: results}. In Sculptor, the fitted surface density model provides a satisfactory match to the data, consistent with equilibrium configurations in simulations. In Fornax, by contrast, the residual power spectrum shows excess low-frequency power, and the filtered $\chi$ map displays ring-like large-scale fluctuations similar to those recovered in our controlled subhalo experiments, making Fornax an especially interesting case for deeper follow-up.}
\label{fig: gaia-test}
\end{figure*}

To this end, we applied a liberal interloper cleaning to both galaxies (Appendix~\ref{app: data-clean}), then fitted an axisymmetric Plummer model atop a constant background to account for residual contamination (Appendix~\ref{app: axi-plummer-fit}). As in our simulations, this provided a good first-order description of these galaxies' surface density. From these fits, we constructed $\chi$ fields and applied our Fourier-based filter to suppress high-frequency, noise-like fluctuations, producing the low-pass maps shown on the rightmost panels of Figure~\ref{fig: gaia-test}.

For Sculptor, no discernible signal emerges, making further baseline tests unnecessary. Possible explanations include the absence of a sufficiently massive subhalo population, the sparse stellar sampling in \gaia, or other perturbations masking subhalo features (e.g. tides or baryonic feedback). 
More specifically, with fewer than $10^{4}$ member stars in the \gaia\ dataset and a parent DM halo mass of $1.58^{+1.01}_{-0.62} \times 10^8~\msun$ \citep{Vitral+25}, no measurable signal is expected according to our simulations. Indeed, the model with $M_{\rm halo} = 3 \times 10^8~\msun$ and $N_{\star} = 10^{4}$ tracers produced no significant features in either the power spectrum or the corresponding filtered $\chi$ maps.
In contrast, Fornax shows a corrugation pattern reminiscent of our simulations: a central underdensity relative to the Plummer fit (blue), surrounded by an overdense annulus near the scale radius (red). For context, we also mark the locations of Fornax's star clusters,\footnote{We use the positions provided in the Local Volume database from \cite{Pace+24}.} which may introduce local features in the filtered $\chi$ map, but are not to be confused with the capture effects discussed earlier in Section~\ref{sec: results}. We note, however, that \gaia\ DR3 is too shallow to recover all of Fornax's clusters (e.g. Fornax~6; \citealt{Wang+19-fornax6}), implying that some captured structures may also remain undetected.

To test robustness, we also fitted alternative density models. A King profile, long advocated for Fornax (e.g. \citealt{Battaglia+06,Penarrubia+09}), provided comparable statistical quality, yet the residual map still showed a ring-like overdensity. The $\alpha\beta\gamma$ family gave an even better fit\footnote{For clarity, we applied the general parametric form of \cite{Zhao96} to the surface density $\Sigma(R)$, although it is more commonly used for the volume density, $\rho(r)$. Here, it serves simply as a convenient parametrization to fit the data.} and significantly reduced the residuals, but only by allowing the inner slope $\gamma$ to be positive,\footnote{In other words, $\displaystyle \lim_{r \to 0} \left[ \diff \rho / \diff r \right] > 0$.} at at $99.8\%$ confidence, in alignment with the arguments from Section~\ref{sssec: baseline}. As discussed, such centrally decreasing densities are restrictive, and for Fornax -- often suggested to host a DM core \citep[e.g.][]{Goerdt+06,Walker&Penarrubia&Penarrubia11,Jardel&Gebhardt12,Sanders&Evans20} -- particularly problematic: DM cores are usually linked to more radial stellar anisotropy \citep{vanderMarel+00,Vitral+24,Vitral+25}, making equilibrium with a centrally decreasing density even harder to sustain \citep{Ciotti&Morganti09}.

We further evaluate other possible sources of the observed pattern. For instance, Fornax’s globular clusters, each with masses of order $10^{5}~\msun$ \citep{deBoer&Fraser16}, are both less massive and less numerous than the subhalos responsible for the perturbations in our models. Any signatures they might produce would therefore be expected to be much weaker than the signal recovered in Figure~\ref{fig: gaia-test}. In fact, the perturbations in Fornax’s \gaia\ data resemble those in our $M_{\rm halo} = 10^{9}~\msun$ model, where subhalos as massive as $10^{7}~\msun$ are present (cf. Figure~\ref{fig: subhalo-masses}), rather than the milder fluctuations seen in the $3 \times 10^{8}~\msun$ case (Figure~\ref{fig: mass-low-pass}).\footnote{As a further check, we also ran a test in which a Fornax-like dSph was initialized without subhalos but with a population of globular clusters following the initial conditions from \cite{Vitral&Boldrini22}. This setup produced neither the gravothermal expansion seen in P25 nor noticeable features in the filtered $\chi$ maps, supporting the arguments above.}
Crowding in \gaia\ data also seems unlikely to explain the observed central depletion. The stellar density in Fornax peaks at $\lesssim 2\times 10^{5}~\mathrm{deg}^{-2}$, whereas crowding effects have only been reported at much higher levels—for instance, in $\omega$ Centauri at $\gtrsim 10^{6}~\mathrm{deg}^{-2}$ \citep{Gaia-omegaCen23}, consistent with figure~7 of \cite{Arenou+18}. These thresholds are therefore nearly an order of magnitude above Fornax's peak density.

Taken together, these analyses indicate that, under favorable conditions, a stellar count of $N_{\star} \sim 10^{5}$ may suffice to recover heating signatures similar to those predicted by our toy models with dark subhalos. 
Still, caution is required before attributing the observed density corrugations in Fornax directly to subhalo interactions. Our simulations are idealized, and it remains unclear whether other processes could generate comparable patterns (e.g. tidal stirring or supernova-driven feedback). Figure~\ref{fig: gaia-test} therefore highlights both the immediate applicability of our framework to existing data and the importance of testing against alternative mechanisms. In this sense, Fornax stands out as a particularly promising target for future follow-up.

\section{Conclusion} \label{sec: conclusion}

In the present work, we employed a simplified set of $N$-body toy-experiments from \cite{Penarrubia+25} to investigate whether interactions of dark matter subhalos and the stellar content of dwarf spheroidal galaxies (dSphs) can induce out-of-equilibrium features that are statistically recoverable from fluctuations in the stellar surface density profile.
These perturbations manifest themselves as density corrugations with complex spatial patterns relative to a baseline density profile with a constant stellar core. In some cases, localized overdensities also emerge, tracing stars temporarily \textit{captured} by individual subhalos (\citealt{Penarrubia+24}). In this paper, we study surface density radial fluctuations in Fourier space -- an approach that not only enables the identification of coherent structures across a wide range of spatial scales, 
but also allows for controlled Poisson-like noise filtering and a flexible reconstruction of the fluctuation field. This provides a robust framework for isolating subhalo-induced signals, hard to detect directly in real space, without requiring \textit{a priori} assumptions about their form.

When dark subhalos are present, we observe peaks in the azimuthally averaged Fourier power spectrum that are well described by a Voigt profile -- that is, the convolution of Gaussian and Lorentzian components -- with a prominent peak typically centred at $2 \pi \, \kappa \, r_{\rm scale} \approx 0.4$, where $r_{\rm scale}$ is the scale radius of the best-fitting axisymmetric Plummer profile (see Eq.~\ref{eq: axi-sd-plummer}). We find that the peak's position, effective width, and normalized flux are highly sensitive to both the halo–subhalo mass model and the number of stellar tracers used to construct the spectrum. In particular, signal strength increases with a larger number of visible tracers, with pronounced detections -- given our halo, subhalo and stellar setups -- occurring at $N_{\star} = 10^{5}$, even for subhalo populations with upper mass limits of $\lesssim 10^{6}~\msun$. Larger subhalo masses lead to stronger signals and more complex corrugated patterns in the projected stellar density field. 

Using archival data from the \gaia\ mission and \hst\ for classical Milky Way dwarf spheroidals, we assessed the prospects for detecting subhalo-induced signals. For Sculptor, the axisymmetric Plummer baseline provides an excellent fit, consistent with equilibrium configurations in simulations. For Fornax, however, the residual power spectrum shows low-frequency structures akin to those recovered in our controlled subhalo experiments. Among the density models tested, only the $\alpha\beta\gamma$ family could fade this signal, but at the cost of introducing a density that decreases towards the centre -- a configuration often viewed as difficult to reconcile with stability, unless restrictive conditions on stellar velocity anisotropy and halo structure are met (see Section~\ref{sssec: baseline}). This makes Fornax a particularly interesting case for follow-up with deeper data. In the coming years, our tests suggest that upcoming wide-field facilities with greater photometric depth -- such as \textit{Euclid}, the \textit{Nancy Grace Roman Space Telescope}, and the \textit{Vera C. Rubin Observatory} -- are expected to recover stellar counts of $N_{\star} \sim 10^{5}$ per dwarf, offering some of the most promising tools for probing subhalo imprints within the Local Group.

Looking forward, several promising directions exist for extending this work. These include studying the impact of different dark matter candidates on the spectral features characterized here, as well as incorporating other dynamical perturbations such as tidal effects from host galaxies, mergers with minor companions, and baryonic processes (see Section~5 of \citealt{Penarrubia+25} for a detailed discussion).
In particular, tidal forces represent a valuable avenue for further investigation, as they can significantly affect the internal structure and stability of dwarf galaxies in both $\Lambda$CDM and modified-gravity frameworks \citep[e.g.][]{Asencio+22}, offering an additional pathway to probe the presence of dark matter.
Future extensions of this work will also benefit from including kinematic information, which provides complementary constraints on the equilibrium state of these systems and on the nature of the stellar overdensities identified in our analysis.\footnote{We plan to explore this aspect in detail in the forthcoming paper of the present series.}
Incorporating these effects will enable more complete and robust constraints on the subhalo mass function. Ultimately, the framework developed here introduces a new and complementary approach to constraining clumpiness of DM in the Local Group dwarf spheroidal galaxies -- the most dark matter-dominated galaxies in the known Universe.

\section*{Acknowledgements}

We thank the anonymous referee for providing insightful comments that improved the clarity of the manuscript.
Eduardo Vitral acknowledges funding from the Royal Society, under the Newton International Fellowship programme (NIF\textbackslash R1\textbackslash 241973).
MGW acknowledges support from the National Science Foundation (NSF) grant AST-2206046. Support for program JWST-AR-02352.001-A was provided by NASA through a grant from the Space Telescope Science Institute, which is operated by the Association of Universities for Research in Astronomy, Inc., under NASA contract NAS 5-03127. This material is based upon work supported by the National Aeronautics and Space Administration under Grant/Agreement No. 80NSSC24K0084 as part of the Roman Large Wide Field Science program funded through ROSES call NNH22ZDA001N-ROMAN.

We have benefited from insightful discussions on Fourier analyses with Shweta Dalal, on the stability of stellar systems with Anna Lisa Varri, on heating mechanisms with Paola Di Matteo, and on the general scientific aspects of this manuscript with Raphaël Errani.

This work has made use of data from the European Space Agency (ESA) mission \gaia\ (\url{https://www.cosmos.esa.int/gaia}), processed by the \gaia\ Data Processing and Analysis Consortium (DPAC, \url{https://www.cosmos.esa.int/web/gaia/dpac/consortium}). Funding for the DPAC has been provided by national institutions, in particular the institutions participating in the \gaia\ Multilateral Agreement.

The following software was used during the writing of this manuscript: {\sc Python} \citep{VanRossum09}, 
{\sc BALRoGO} \citep{Vitral21},
{\sc Scipy} \citep{Jones+01},
{\sc Numpy} \citep{vanderWalt11},
{\sc Matplotlib} \citep{Hunter07},
and the Scientific color maps from \cite*{Crameri+20}.
\section*{Data Availability}

The data that support the plots within this paper and other findings of this study are available from the corresponding author upon reasonable request.



\bibliographystyle{mnras}
\bibliography{src} 




\appendix


\section{Analytical handling of the Plummer density profile}

\subsection{Plummer mixture model}  \label{app: axi-plummer-fit}

We present the analytical formalism for a mixture model comprising $N$ axisymmetric Plummer components \citep{Plummer1911}, along with a constant-density background representing interlopers. For a single Plummer component in projected geometry, the axisymmetric surface density profile is given by:
\begin{equation} \label{eq: axi-sd-plummer}
\begin{split}
    \Sigma_{\rm sys}(R, \xi) = \frac{N_{\infty}}{\pi r_{\rm scale}^{2} (1 - \epsilon)} \,
    \left[1 + \frac{R^{2}}{r_{\rm scale}^{2}} \left(\cos^{2}\xi + \frac{\sin^{2}\xi}{(1-\epsilon)^{2}}\right) \right]^{-2} \ ,
\end{split}
\end{equation}
where $N_{\infty}$ is the total number of tracers per component, and the ellipticity is defined as $\epsilon \equiv 1 - b/a$, where $a$ and $b$ are the major and minor axis lengths, respectively. For consistency with the remainder of this work, we define $r_{\rm scale} \equiv a$. In a multi-component Plummer mixture, each component has its own definition of $R$ and $\xi$, such that
\begin{flalign}
    R &= \sqrt{x^{2} + y^{2}}   \ , \\
    \xi &=  {\rm arctan2}(y, x) - \phi  \ .
\end{flalign}

In the relation above, the Cartesian coordinates $(x, y)$ are centered on the component’s centre $(x_0, y_0)$, and $\phi$ indicates the projected angular coordinate over the major axis. For datasets already in Cartesian coordinates, this implies:
\begin{flalign} \label{eq: cart-coords}
    x &= x_{\rm true} - x_{0}   \ , \\
    y &= y_{\rm true} - y_{0}  \ ,
\end{flalign}
where $(x_{\rm true}, y_{\rm true})$ are the original coordinates. For data projected on the sky, the appropriate Cartesian transformation is given by, e.g. \citealt{GaiaCollaborationHelmi+18}:
\begin{flalign}
    x &= \cos{\delta} \sin{(\alpha - \alpha_{0})}   \ , \\
    y &= \sin{\delta} \cos{\delta_{0}} - \cos{\delta} \sin{\delta_{0}} \cos{(\alpha - \alpha_{0})} \ ,
    \label{eq: proj-coords}
\end{flalign}
where $(\alpha, \delta)$ are the celestial coordinates, and $(\alpha_0, \delta_0)$ denote the centre of the corresponding Plummer component in this coordinate frame.

In this work, we apply the formalism to numerical data representing a dwarf galaxy free from contamination by interlopers. In realistic scenarios, however, a contribution from foreground and background stars is often present. If these contaminants are assumed to be uniformly distributed across the sky, their surface density can be described by:
\begin{equation}
    \Sigma_{\rm ilop} = N_{\rm ilop} / \mathcal{S}_{\rm data} \ ,
\end{equation}
where $N_{\rm ilop}$ is the number of contaminant stars and $\mathcal{S}_{\rm data}$ is the surface area over which the data are defined.

The likelihood of the full mixture model is then given by:
\begin{equation}
    \mathcal{L} \equiv \prod_{\rm j} \frac{{\diff \probP} }{\diff \mathcal{S}} = \prod_{\rm j} \left( \Sigma_{\rm ilop} / N_{\rm tot} + \sum_{\rm i}^{N} \Sigma_{\mathrm{sys}, \, i} / N_{\rm tot} \right) \ ,
\end{equation}
where $N_{\rm tot} = N_{\rm ilop} + \sum_{\rm i}^{N} N_{\infty, i}$ is the total number of tracers, and $j$ indexes individual data points. This formalism includes a few simplifying assumptions: first, that field stars (interlopers) are uniformly distributed across the sky, which may not always hold in real observations; and second, that the data coverage $\mathcal{S}_{\rm data}$ fully encloses all tracers.
These assumptions hold exactly in our simulations, which use uncontaminated data (i.e., $N_{\rm ilop} = 0$) and include all stars in the fit. In real datasets, such assumptions are valid when interloper contamination is either negligible or spatially uniform, and when the survey footprint captures most member stars. For our \gaia-based tests on Sculptor and Fornax, we verified that interlopers are approximately uniformly distributed across the field, and that the $5\fdg0$ cone search employed in our analysis sufficiently encompasses the majority of galaxy members. Under these conditions, we fit the model by maximizing the log-likelihood, $\ln \mathcal{L}$.

\subsection{Stellar density mapping using a square grid}    \label{app: count-cell}

One of the steps outlined in Section~\ref{sec: methods} involves computing the number of stars contained within each cell of a square grid. In the simplified case considered here -- namely, a single Plummer component with interlopers -- this corresponds to evaluating the following integral:
\begin{equation} \label{eq: cell-count}
    N_{\rm ij} = \int_{x_{i, \rm min}}^{x_{i, \rm max}} \int_{y_{j, \rm min}}^{y_{j, \rm max}} \Sigma_{\rm sys} \, \diff x \diff y  \   +  \Sigma_{\rm ilop} \, \Delta x_{i} \, \Delta y_{j},
\end{equation}
where $N_{\rm ij}$ is the number of stars within the grid cell indexed by $\rm ij$, and $x_{i, \rm min}$, $x_{i, \rm max}$, $y_{j, \rm min}$, and $y_{j, \rm max}$ define the boundaries of the cell along the $x$ and $y$ axes. Here, $\Sigma_{\rm sys}$ denotes the Plummer surface density profile, as defined in eqs.~\ref{eq: axi-sd-plummer} -- \ref{eq: proj-coords} and $\Sigma_{\rm ilop}$ is the constant interloper background density.
For simplicity, we assume the grid is centered on the Plummer component and aligned with its major and minor axes.
If not, the coordinate system (i.e., the $(x, y)$ coordinates from eqs.~\ref{eq: cart-coords} -- \ref{eq: proj-coords}) is rotated by an angle $\varphi$ so that the $x$-axis aligns with the major axis:
\begin{equation}
    \begin{pmatrix}
        x_{\rm rot} \\
        y_{\rm rot} \\
    \end{pmatrix} = 
    \begin{pmatrix}
        \cos\varphi & -\sin\varphi \\
        \sin\varphi & \cos\varphi \\
    \end{pmatrix}
    \begin{pmatrix}
        x \\
        y \\
    \end{pmatrix} \ .
\end{equation}
Once the axes and centre have been aligned, we construct a square grid with uniform resolution $\delta_{\rm grid}$ along each Cartesian direction. We define $\delta_{\rm grid}$ as a function of the Plummer major axis, $r_{\rm scale}$, and the size of the two-dimensional square box $\Delta$ used in our analysis. Specifically, we set
\begin{equation}
\delta_{\rm grid} = \Delta \times \round{(64 \, \Delta / r_{\rm scale})}^{-1} \ ,
\end{equation}
where $\round{x}$ denotes the nearest integer to $x$. In our case, we adopt $\Delta = 20$~kpc, as density fluctuations beyond this scale are unlikely to be reliably measured in observations, or to be physically meaningful for our science case.

We then normalize the $(x, y)$ coordinates by the respective major and minor axis lengths of the Plummer profile (i.e. $r_{\rm scale}$ and $r_{\rm minor} = (1 - \epsilon) \, r_{\rm scale}$, respectively) to eliminate grid asymmetries. For notational simplicity, we denote the Cartesian boundaries of grid cell $\rm ij$ as $(x_{\rm min}, x_{\rm max}, y_{\rm min}, y_{\rm max})$. Within this framework, the stellar count associated with the Plummer component in each cell, $N_{\rm ij, \, sys}$ (i.e. the integral in eq.~\ref{eq: cell-count}), can be computed analytically as
\begin{equation}
    N_{\rm ij, \, sys} = \frac{1}{2D} \left( T_{1} + T_{2} + T_{3} + T_{4}\right) \ ,
\end{equation}
where we have introduced the terms
\begin{flalign}
    D &= \sqrt{\left(x_{\rm max}^2+1\right) \left(x_{\rm min}^2+1\right)
   \left(y_{\rm max}^2+1\right) \left(y_{\rm min}^2+1\right)}   ,    \\[20pt]
   T_{1} &= x_{\rm max}
   \sqrt{\left(x_{\rm min}^2+1\right) \left(y_{\rm max}^2+1\right)
   \left(y_{\rm min}^2+1\right)}  \nonumber \\
   & \times \left[ \tan
   ^{-1}\left(\frac{y_{\rm max}}{\sqrt{x_{\rm max}^2+1}}\right) - \tan
   ^{-1}\left(\frac{y_{\rm min}}{\sqrt{x_{\rm max}^2+1}}\right) \right] ,   \\[20pt]
   T_{2} &= x_{\rm min}
   \sqrt{\left(x_{\rm max}^2+1\right) \left(y_{\rm max}^2+1\right)
   \left(y_{\rm min}^2+1\right)}  \nonumber \\
   & \times \left[ \tan
   ^{-1}\left(\frac{y_{\rm min}}{\sqrt{x_{\rm min}^2+1}}\right) - \tan
   ^{-1}\left(\frac{y_{\rm max}}{\sqrt{x_{\rm min}^2+1}}\right) \right] ,   \\[20pt]
   T_{3} &= y_{\rm max}
   \sqrt{\left(y_{\rm min}^2+1\right) \left(x_{\rm max}^2+1\right)
   \left(x_{\rm min}^2+1\right)}  \nonumber \\
   & \times \left[ \tan
   ^{-1}\left(\frac{x_{\rm max}}{\sqrt{y_{\rm max}^2+1}}\right) - \tan
   ^{-1}\left(\frac{x_{\rm min}}{\sqrt{y_{\rm max}^2+1}}\right) \right] ,   \\[20pt]
   T_{4} &= y_{\rm min}
   \sqrt{\left(y_{\rm max}^2+1\right) \left(x_{\rm max}^2+1\right)
   \left(x_{\rm min}^2+1\right)}  \nonumber \\
   & \times \left[ \tan
   ^{-1}\left(\frac{x_{\rm min}}{\sqrt{y_{\rm min}^2+1}}\right) - \tan
   ^{-1}\left(\frac{x_{\rm max}}{\sqrt{y_{\rm min}^2+1}}\right) \right] .
\end{flalign}

We thus refer to this formalism whenever estimating the number of stars from a Plummer profile within the cell of a square grid.

\section{Handling of the power spectrum} \label{app: voigt-filter}

\begin{figure}
\centering
\includegraphics[width=\hsize]{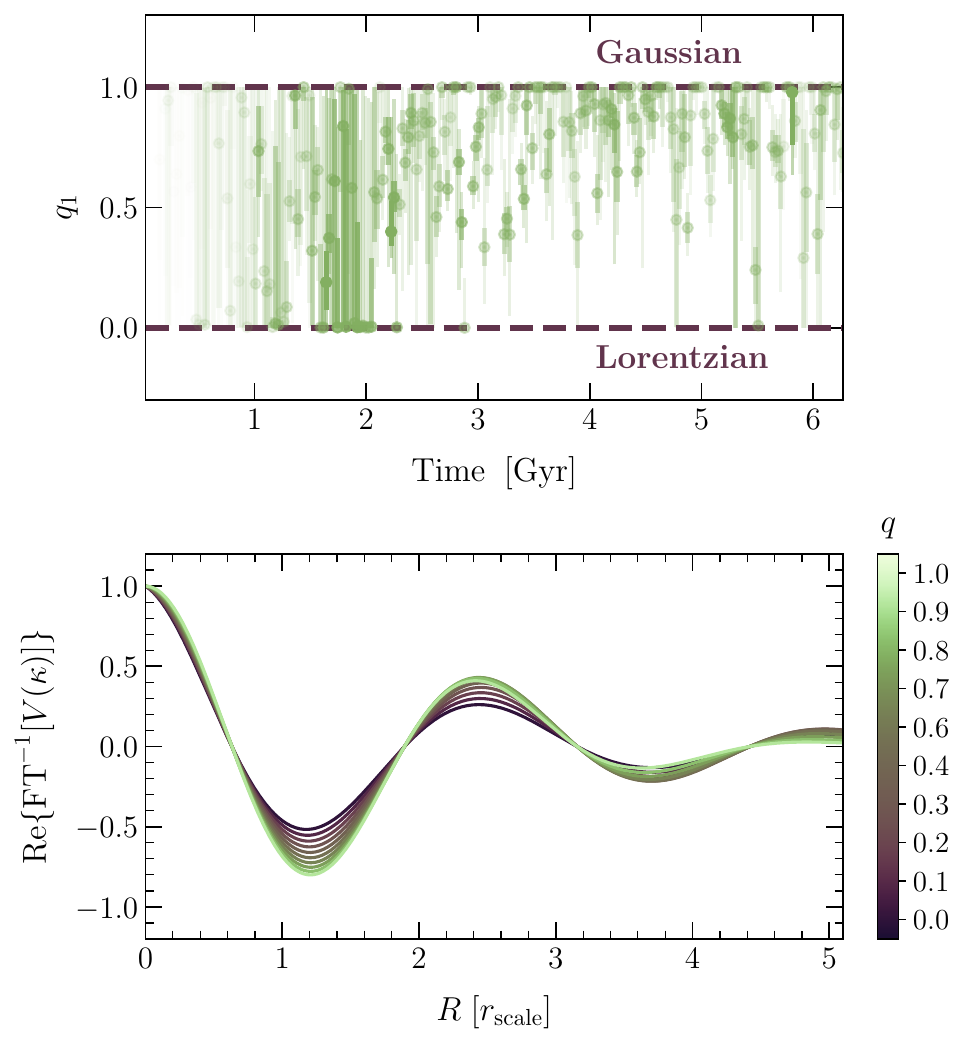}
\caption{\textit{Voigt mixture parameter:}
The top row shows the time evolution of the fitted $q_{1}$ mixture parameter from our Voigt fits using Eq.~\ref{eq: power-spectrum}, for the simulation with $M_{\rm halo} = 10^{9}~\msun$ and $N_{\star} = 10^{5}$ stellar particles. Green dots indicate the median values at each snapshot, with error bars representing the 16th–84th percentile range. Point transparency is scaled inversely with the normalized flux, $f_{1} / \mathcal{C}$, to emphasize more confidently measured values. For reference, we plot the pure Gaussian ($q = 1$) and Lorentzian ($q = 0$) limits of the Voigt profile as dashed purple lines, illustrating that many well-constrained $q_{1}$ values lie between these extremes -- justifying the need for a flexible Voigt parametrization.
The bottom row, for pedagogical purposes, displays the Voigt profile from Eq.~\ref{eq: voigt_real} evaluated at the $\mu_{1}$ and $\delta_{1}$ values corresponding to the 4~Gyr snapshot of this simulation, while varying the mixture parameter from $q = 0$ to a $q = 1$. The resulting peak shapes closely resemble the structures seen in the filtered $\chi$ maps, reinforcing the interpretability of our spectral decomposition.}
\label{fig: q-vs-t}
\end{figure}

\begin{figure*}
\centering
\includegraphics[width=0.75\hsize]{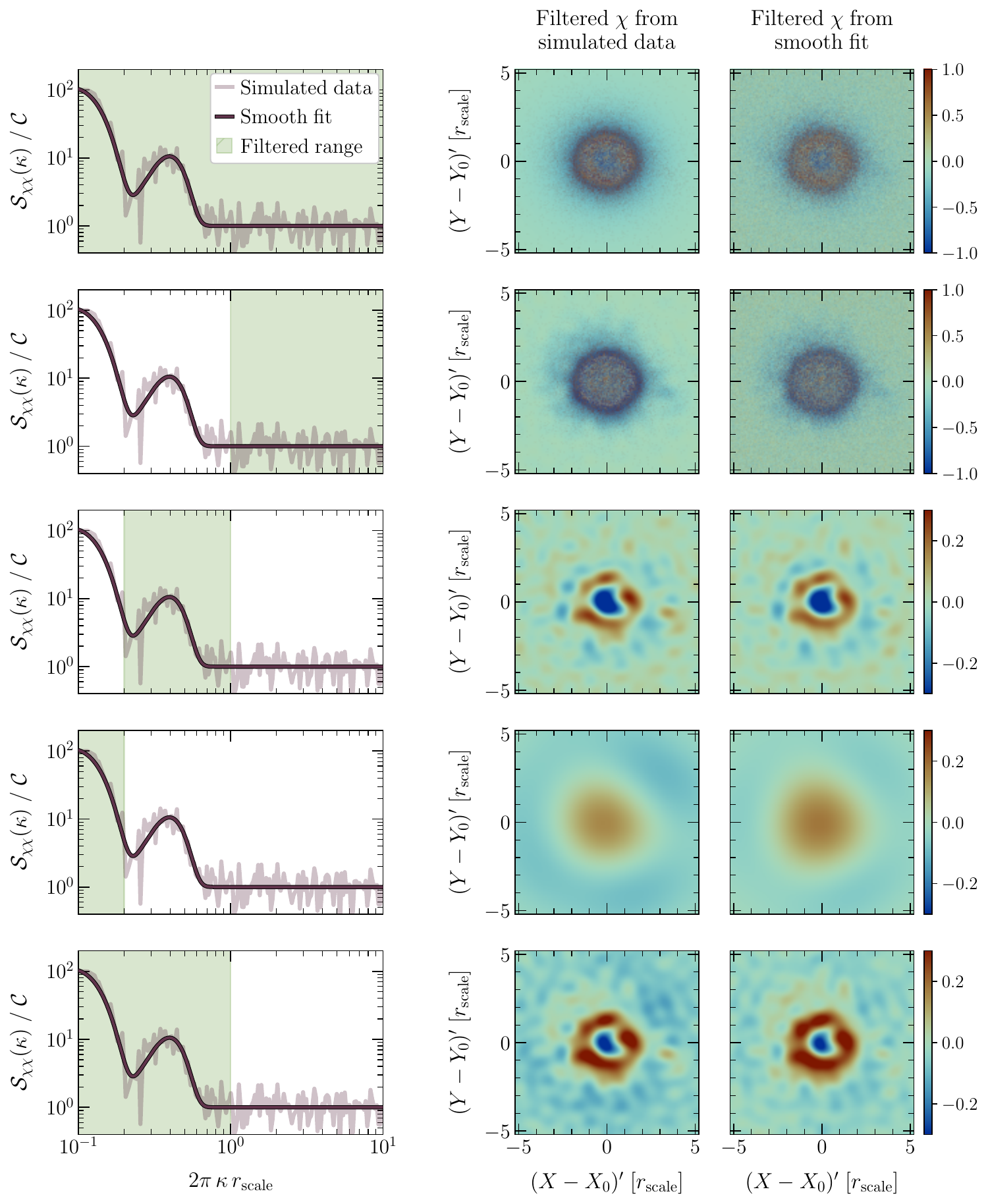}
\caption{\textit{Filtered $\chi$ maps:}
Illustration of the connection between selected regions of the Fourier power spectrum and their corresponding filtered $\chi$ maps, shown for a representative snapshot at $4$~Gyr from the simulation with $M_{\rm halo} = 10^{9}~\msun$ and $N_{\star} = 10^{5}$ stellar particles. Left panels display both simulated and fitted power spectrum (light and dark purple, respectively), with the active filtering range highlighted in green. Middle and right panels show the resulting $\chi$ maps constructed from the respective simulated and fitted spatial frequency intervals (both use the simulated phase information), with color limits adjusted at each row to improve visualization. Each row corresponds to a different filtered range. The results indicate that the most informative signals emerge from low spatial frequencies, at $2 \pi \, \kappa \, r_{\rm scale} < 1$, where the two Voigt components from Eq.~\ref{eq: power-spectrum} reside. In particular, the primary peak near $2 \pi \, \kappa \, r_{\rm scale} \approx 0.4$ produces the most intricate and spatially structured $\chi$ features, while the secondary peak plays a predominantly smoothing role, offering little added value for the analysis.}
\label{fig: filtered-chi}
\end{figure*}

This appendix provides additional context for the use of a Voigt profile, as well as the filtering choices and cutoff frequency adopted in Section~\ref{ssec: low-pass-filter}. To illustrate our approach, we analyze a representative snapshot at 4~Gyr from the simulation with $M_{\rm halo} = 10^{9}~\msun$ and $N_{\star} = 10^{5}$, where subhalo-induced density fluctuations are clearly visible (see, e.g., Figure~\ref{fig: power-spectrum}).

\subsection{Choice of a Voigt profile} \label{app: voigt-mixture}

Voigt profiles are inherently more flexible than simple Gaussian or Lorentzian functions, combining both forms to accommodate a broader range of spectral shapes. This appendix provides further motivation for their use in fitting the power spectra of surface density fluctuations.

To illustrate this need, the top row of Figure~\ref{fig: q-vs-t} shows the evolution of the mixture parameter $q_{1}$ from the Voigt profile defined in Eq.~\ref{eq: power-spectrum}, evaluated across snapshots of our fiducial simulation. The parameter abruptly interpolates between a pure Lorentzian ($q = 0$) and a pure Gaussian ($q = 1$), and the figure demonstrates that many snapshots yield well-constrained values of $q_{1}$ that lie between these extremes. This underscores the importance of a profile that can flexibly adapt to different spectral shapes.

To further illustrate the effect of $q$, the bottom row of the same Figure shows the corresponding real-space density fluctuation associated with the primary Voigt peak, as given by Eq.~\ref{eq: voigt_real}. These curves are sampled at the best-fit values of $\mu_{1}$ and $\delta_{1}$ for the 4~Gyr snapshot. The results reveal that this parameter regulates the amplitude of the density fluctuations (along with the associated flux $f_{1}$), yielding more pronounced features in the Gaussian-dominated regime at smaller spatial scales, and weaker fluctuations in this same regime at larger scales, where the Lorentzian scenario becomes more influential.

\subsection{Choice of filtering ranges} \label{app: chi-reconstruct}

To better contextualize the filtering choices and cutoff frequency adopted in Section~\ref{ssec: low-pass-filter}, we construct Figure~\ref{fig: filtered-chi}, which presents a series of filtered $\chi$ maps derived from distinct frequency intervals of the power spectrum. In each row, the left panel shows the observed and fitted spectrum with the active filtering range shaded in green; the middle panel displays the corresponding filtered $\chi$ map from the raw simulated data, and the right panel shows the same from the smooth Voigt mixture, although the phase information is still sourced from the raw simulation. This decomposition reveals that the dominant contributions from subhalo-induced signals arise at low spatial frequencies, $2 \pi \, \kappa \, r_{\rm scale} < 1$, coinciding with the Voigt peaks described in Eq.~\ref{eq: power-spectrum}. In particular, the primary peak near $2 \pi \, \kappa \, r_{\rm scale} \approx 0.4$ generates the most structured and informative $\chi$ features, whereas the lower-frequency secondary peak contributes primarily to mild smoothing, with limited utility for distinguishing between halo–subhalo mass models. 

These results support the filtering strategy adopted in the main analysis, which effectively isolates the spatial scales most sensitive to subhalo-induced perturbations, and also validate the use of a Voigt mixture to represent the corresponding power spectrum. Finally, it also serves as motivation for our Fourier formalism outlined in Section~\ref{sec: methods}, since the fluctuations in the upper row $\chi$ map are difficultly observable without appropriate filtering.

\section{Handling of \gaia\ data} \label{app: data-clean}

In Section~\ref{sssec: test-gaia}, we use \gaia\ DR3 data to apply our formalism to observed datasets of classical dwarfs. While Section~\ref{ssec: expectations-stellar-counts} considered the full catalog within a $5\fdg0$ cone around each galaxy, in Section~\ref{sssec: test-gaia} we applied a deliberately liberal cleaning procedure over the same area. This reduced the ratio of interlopers to dwarf stars, yet still produced a member sample of the same order of magnitude as previously estimated. The present section details the applied cuts for reproducibility.

\begin{figure}
\centering
\includegraphics[width=\hsize]{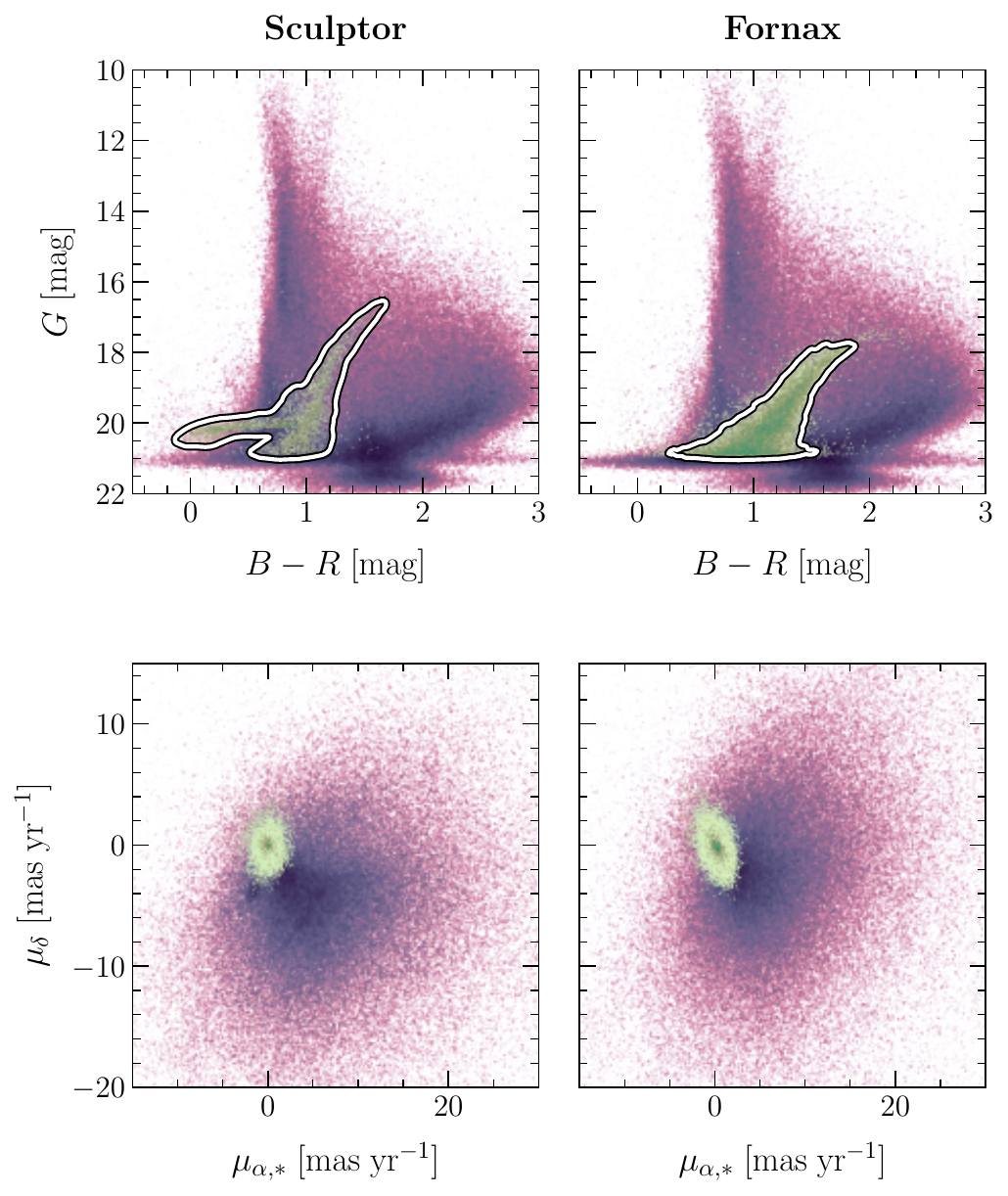}
\caption{\textit{Selection function:}
Illustration of our liberal cleaning of \gaia\ DR3 data, aimed at reducing interlopers while retaining a large fraction of dwarf members. The left panels correspond to Sculptor and the right panels to Fornax. In all panels, the full set of stars with available parameters is shown in purple, with darker tones marking higher-density regions.
The upper panels display the color–magnitude diagram. Green points mark stars with combined spatial and proper-motion membership probability $>0.5$, while the white curve traces a 2.5-$\sigma$ Kernel Density Estimate contour (bandwidth equal to half the \protect\citealt{Silverman86} rule), using those same probabilities as weights.
The lower panels show the proper-motion plane. Here, green points indicate stars with proper-motion membership probability $>0.5$.
The cleaned sample used in Section~\ref{sssec: test-gaia} consists of the stars inside the white CMD mask, the proper-motion–selected green points, and all stars with positional information only. We note that the conclusions of Section~\ref{sssec: test-gaia} remain unchanged if interloper-related cuts are omitted, though the signal-to-noise is reduced.}
\label{fig: gaia-selection}
\end{figure}

The procedure began by fitting the uncleaned data's surface density with an axisymmetric Plummer profile. In parallel, we modeled the joint proper-motion distribution of dwarf members and interlopers using the Gaussian plus asymmetric Pearson~VII mixture introduced in \cite{Vitral21}. Together, these models provided a membership probability for each star, combining spatial and kinematic information; stars without proper-motion measurements were assigned \texttt{NaN} probabilities. For reference, those with probability $>0.5$ -- i.e. likely members -- are shown as green points in the upper panels of Figure~\ref{fig: gaia-selection}.
These membership probabilities were then used as weights in a Kernel Density Estimation (KDE) of the color–magnitude diagram (CMD), once again following the procedure suggested in \cite{Vitral21}.\footnote{We adopted a 2.5-$\sigma$ KDE contour with bandwidth set to half the \protect\cite{Silverman86} rule, using the membership probabilities as weights.} Based on this KDE, we defined a broad mask in CMD space (white lines in Figure~\ref{fig: gaia-selection}), rejecting stars outside the mask while retaining those without CMD information. This step yielded a first cleaned set of members.

Independently, within the uncleaned sample, we also selected stars with proper-motion membership probability $>0.5$, identifying them as likely members (green points in the lower panels of Figure~\ref{fig: gaia-selection}) and forming a second member set. We then combined both selections -- the CMD-based and the proper-motion–based filters -- into a single cleaned dataset, while retaining stars without such information (i.e. those for which only positions are available).

By construction, this dataset does not introduce spatial incompleteness and is therefore suitable for our surface-density analysis in Section~\ref{sssec: test-gaia}. We note that the conclusions of that section remain robust even when interloper-related cuts are omitted, although in that case the transition from high-density dwarf members to the interloper background occurs at smaller radii, which eventually obscures lower-frequency signals in our Fourier spectral analysis.


\bsp	
\label{lastpage}
\end{document}